\documentclass[a4paper,11pt]{article}
\usepackage{jheppub}
\usepackage{amsfonts}
\usepackage{mathrsfs}
\usepackage{graphicx}
\usepackage{amsmath}
\usepackage{amssymb}
\usepackage{bbm}
\usepackage{subfigure}
\usepackage{caption}
\usepackage{float}
\usepackage{epsfig}
\usepackage{graphicx}
\usepackage{tikz}
\usepackage{color}
\usepackage{physics}
\usepackage{indentfirst}
\usepackage{hyperref}
\usepackage{setspace}
\usepackage{braket}
\usepackage{slashed}
\usepackage{multirow}
\usepackage{diagbox}
\usepackage{cancel}
\usepackage{caption}
\usepackage{array}
\usepackage{ulem}
\usepackage{bibentry}
\newcommand{\dslah}{\slash \! \! \! }

\newcommand{\mi}{\mathrm{i}}

\allowdisplaybreaks[4]

\arxivnumber{2512.22019}

\title{\boldmath Mass Spectra of $\Lambda_Q\bar{\Sigma}_Q$ Hexaquark States in QCD Sum Rules}

\author[a]{Xuan-Heng Zhang,}
\author[a,b]{Cong-Feng Qiao}
\affiliation[a]{School of Physical Sciences, University of Chinese Academy of Sciences,\\
YuQuan Road 19A, Beijing 100049, China}
\affiliation[b]{International Centre for Theoretical Physics Asia-Pacific, University of Chinese Academy of Sciences, Beijing 100190, China}
\emailAdd{qiaocf@ucas.ac.cn}

\abstract{Recently, the BESIII Collaboration indicate that no $\Lambda_c\bar{\Sigma}_c$ bound-state with a mass near threshold in the range $4715$--$4735~\mathrm{MeV}$ was observed. In order to determine the plausible mass region of the states in this structure, we calculate the mass spectrum of the $\Lambda_c\bar{\Sigma}_c$ configuration with the method of QCD sum rules. Two linearly independent interpolating currents are constructed, and contributions from nonperturbative condensates up to dimension 12 are included in the numerical results. Consequently, we obtain the masses of the candidate states with quantum numbers $J^P = 0^-,\,0^+,\,1^-,\,1^+$. Our results show that the central values of the $\Lambda_c\bar{\Sigma}_c$ ground-state masses lie around the $5.8~\mathrm{GeV}$ region, which do not support them as bound states and consistent with the findings reported by the BESIII Collaboration. Furthermore, we compute the mass spectrum of the $\Lambda_b\bar{\Sigma}_b$ states with quantum numbers $J^P = 0^-,\,0^+,\,1^-,\,1^+$, which could be served as hidden-bottom candidates in the experimental detecting.}

\begin{document}
\maketitle
\flushbottom

\newpage

\section{Introduction}
In the 1960s, in order to explain the newly discovered particles observed at high-energy colliders, Gell-Mann \cite{Gell-Mann:1964ewy} and Zweig \cite{Zweig:1964jf}  independently proposed the quark model (QM), marking the beginning of human exploration of the strong interaction. In the quark model, the allowed hadronic states include mesons, which are bound states of a quark and an antiquark ($q\bar q$) , and baryons, which are composed of three quarks ($qqq$). However, the existence of more complex hadronic configurations is not forbidden by QCD,  examples include tetraquark states, pentaquark states, hexaquark states, glueballs, and hybrid states, which are named as exotic states.  After the development of QCD, three additional heavy quarks predicted by the quark model were experimentally discovered, namely the charm quark $(c)$ \cite{E598:1974sol, SLAC-SP-017:1974ind} , bottom quark $(b)$ \cite{E288:1977efs} , and top quark($t$) \cite{CDF:1995wbb} , marking the beginning of the exploration of heavy-flavor physics. Consequently, the study of heavy-flavor exotic states was put on the agenda, which could enrich our understanding of hadronic structures and the non-perturbative effects of QCD.

In 2003, the Belle II Collaboration discovered a resonance $X(3872)$ at the $D^0\bar{D}^{*0}$ threshold \cite{Belle:2003nnu},  which is currently interpreted theoretically as a tetraquark state. This marked the beginning of both theoretical and experimental studies on the exotic states. In 2015, the $P_c(4380)$ and $P_c(4450)$ resonances reported by the LHCb Collaboration \cite{LHCb:2015yax} were interpreted as  pentaquark states, representing the first discovery of pentaquark structures. Over the past two decades, numerous theoretical explanations have been proposed for these resonances \cite{Pakvasa:2003ea,Matheus:2006xi,Lee:2008uy,Chen:2013pya,Wang:2015epa}, most of which suggested that they are hidden-charm multiquark states \cite{Pakvasa:2003ea,Chen:2013pya,Wang:2015epa} containing heavy quarks. Meanwhile, many additional $XYZ$ states that may correspond to hidden-charm multiquark configurations were observed experimentally, such as $Y(4260)$ \cite{BaBar:2005hhc} , $Y(4660)$ \cite{Belle:2007umv} , $Z_c(3900)$ \cite{BESIII:2013ris} , $Z_c(4200)$ \cite{Belle:2014nuw} and others, which sparked a surge of interest in the study of heavy-flavor exotic states.

Compared to tetraquark and pentaquark states, research on hexaquark states is currently limited. The most widely accepted hexaquark state today is the deuteron, which is a $J^P=1^+$ state of a proton and a neutron with a di-baryon structure \cite{Weinberg:1962hj}.  Currently, there is insufficient experimental evidence to confirm the observation of a new hexaquark state \cite{BaBar:2018hpv}.  Theoretical studies of hexaquark states can be traced back to 1949, when E.Fermi and C.N. Yang proposed that a $p\bar{N}$ state could be used to explain the structure of pion \cite{Fermi:1949voc}. Although this configuration of the pion was later replaced by the quark model, it marked the beginning of the exploration of hexaquark states. Since then, many theoretical predictions of hexaquark states have been made \cite{Jaffe:1976yi,Mulders:1980vx,Balachandran:1983dj,Ikeda:2007nz,Shanahan:2011su,Clement:2016vnl}.   Hexaquark states formed by a bound baryon-antibaryon pair may possess greater stability than those with a di-baryon structure. Such states are referred to as baryoniums. Currently, there are many candidates for baryonium states, including both light-flavor \cite{BES:2003aic,BESIII:2023vvr,BESIII:2023kgz} and heavy-flavor \cite{BaBar:2005hhc,Belle:2007umv} structures. For heavy-flavor baryonium states, numerous theoretical explanations have been proposed, such as explaining the production and decay of the $Y(4260)$ resonance with a $\Lambda_c \bar{\Lambda}_c$ structure \cite{Qiao:2005av,Qiao:2007ce}.

In the exploration of hadronic structures, many theoretical methods have been proposed. Among them, the QCD sum rules (QCDSR) method, introduced in 1979 \cite{Shifman:1978bx,Shifman:1978by}, provides an effective framework that incorporates both perturbative and nonperturbative contributions and yields analytic predictions for hadron spectra. QCDSR were first applied to explain the mass and decay constant of the $\rho$ meson \cite{Shifman:1978by} , with results that are in excellent agreement with the current Particle Data Group (PDG) values \cite{ParticleDataGroup:2024cfk} .  QCDSR can similarly be applied to calculate the mass spectra of exotic states \cite{Colangelo:2000dp,Wang:2025sic} . For instance, multiquark configuration with hidden charm have been constructed to account for the mass of $X(3872)$ , $P_c(4380)$ and $P_c(4450)$, etc. \cite{Nielsen:2009uh}   . For hexaquark states, numerous studies based on QCDSR have also been conducted, including investigations of light baryonium states \cite{Wan:2021vny,Wang:2006sna, Zhang:2024ulk}, heavy baryonium states \cite{Chen:2016ymy,Wan:2019ake, Wang:2021qmn,Wang:2021pua}, and compact hexaquark configurations \cite{Zhang:2025vqg, Wang:2022jvk,Wang:2017sto}. These include studies on hidden-charm baryonium states: In Ref.~\cite{Chen:2016ymy} , the structure of hidden-charm hexaquark states was systematically studied, and the mass of these states was found to be around 5.0 GeV. Refs.~\cite{Wan:2019ake,Wang:2021qmn} both calculated the mass of $\Lambda_c \bar{\Lambda}_c$, with results also around 5.0 GeV. Furthermore, Ref.~\cite{Wan:2019ake} predicted the mass of $\Lambda_b \bar{\Lambda}_b$, while Ref.~\cite{Wang:2021qmn} provided the mass of the $\Lambda_c \Lambda_c$ state with a di-baryon structure, concluding that the central value is around 5.11 GeV. These results can be considered as byproducts of the study on hidden-charm baryonium states.

In the hidden-charm baryonium states, the mass of the $\Lambda_c\bar{\Lambda}_c$ structure could be close to that of $\Lambda_c\bar{\Sigma}_c$, but the theoretical studies on this topic are relatively scarce. Therefore, evaluating the mass of the $\Lambda_c\bar{\Sigma}_c$ state is necessary. Theoretically, a study via  one-boson-exchange potential and Bethe-Salpeter equation method suggested the existence of a hidden-charm $\Lambda_c\bar{\Sigma}_c$ bound state with quantum numbers \((I,S) = (1,0)\) and a mass in the near threshold 4.7--4.8 GeV region \cite{Dong:2021juy}. However, recently, the BESIII Collaboration conducted a search for a possible $\Lambda_c\bar{\Sigma}_c$ bound-state structure near threshold on the BESIII detector and the BEPCII collider \cite{BESIII:2025zgc}. Their results indicated that no such structure with a mass in the range $4715$--$4735~\mathrm{MeV}$ can be observed. The QCDSR calculations for the $\Lambda_c \bar{\Lambda}_c$ system suggest that its mass could exceed 5.0 GeV \cite{Wan:2019ake} . Therefore, it can be conjectured that the mass of $\Lambda_c \bar{\Sigma}_c$ should be even larger. To determine the mass range of the $\Lambda_c \bar{\Sigma}_c$ state, in this work, the masses of ground $\Lambda_c \bar{\Sigma}_c$ states are calculated with QCDSR. Additionally, the masses of the ground-state $\Lambda_b \bar{\Sigma}_b$ are also calculated, which may be observed in future experiments. The structure of this paper is organized as follows: in Sect.\ref{form}, the theoretical framework of QCDSR is briefly introduced and  the fundamental formulas used in our calculations are presented. In Sect.\ref{na}, numerical analyses and results. are provided. Section \ref{dm} discusses the possible decay modes of $\Lambda_Q \bar{\Sigma}_Q$, where $Q=c,b$. Finally, in Sect.\ref{dc}, our results are compared with experimental observations and our findings are summarized.

\section{\label{form}Formalism}
\subsection{Choices of the Currents}
To calculate the mass spectrum of $\Lambda_Q \bar{\Sigma}_Q$ in the framework of QCDSR, it is essential to first select appropriate hadronic interpolating currents. There are two independent interpolating currents for the baryon octet, and the other interpolating current structures can be obtained by linear combinations of them through Fierz transformations \cite{Chen:2008qv}. In our calculation, the masses of quarks $u$ and $d$ are rather smaller than that of heavy quarks, so we take the limit $m_u = m_d \to 0$, which simplifies the structure of the interpolating currents. The two simplified interpolating currents we have chosen are  \cite{Chung:1981cc,Chung:1981wm}
\begin{equation}\label{eq:eta1}
	\text{Type-I:}\quad \eta_{\text{I},\mathscr{B}}(x) =  \varepsilon_{abc}\left[q_a^{i T}(x) \mathcal{C} q_b^j(x)\right] \gamma_5 q_c^k(x)\ ,
\end{equation}
\begin{equation}\label{eq:eta2}
	\text{Type-II:}\quad \eta_{\text{II},\mathscr{B}}(x) =  \varepsilon_{abc}\left[q_a^{i T}(x) \mathcal{C} \gamma_5 q_b^j(x)\right] q_c^k(x)\ ,
\end{equation}
where for $\Lambda_Q, \Sigma_Q$, the indices $(i,j,k)$ take the following values: $(u,d,Q)$ for $\Lambda_Q$, and $(u,Q,d)$ for $\Sigma_Q$. $a,b,c$ are the color indices. $\mathscr{B}$ denotes an arbitrary baryon.

For the baryon-antibaryon type baryonium states $\mathscr{B}\bar{\mathscr{B^{\prime}}}$, the corresponding interpolating current structures are given by
\begin{equation}\label{eq:jcurrent}
	j_{(\mu)}(x)=\bar{\eta}_{\mathscr{B}^{\prime}}\Gamma_{(\mu)}\eta_{\mathscr{B}}(x),
\end{equation}
where $\Gamma_{(\mu)}=\mi\gamma_5,\mathbbm{1},\gamma_{\mu},\gamma_{\mu}\gamma_{5}$ are corresponding to the quantum numbers of the ground states $J^{P} = 0^{-}, 0^{+}, 1^{-}, 1^{+}$. In this work, the two baryonic interpolating currents $\bar{\eta}_{\mathscr{B^{\prime}}}(x)$ and $\eta_{\mathscr{B}}(x)$ for each state are  chosen uniformly from either Eqs.~\eqref{eq:eta1} or \eqref{eq:eta2}. Thus, for each baryonic interpolating current, there exist four possible structures, corresponding to four different quantum numbers $J^{P}$.

\subsection{2-point Correlation Functions}
After selecting the interpolating currents in  Eq.~\eqref{eq:jcurrent}, the two-point correlation functions can be calculated, which are defined as
\begin{equation}\label{eq:Pi0}
	\Pi(q^2) = \mi \int \dd^4 x \, \mathrm{e}^{\mi q \cdot x} 
	\bra{\Omega}\mathbb{T}\{j(x), j^{\dagger}(0)\}\ket{\Omega},
\end{equation}
\begin{equation}\label{eq:Pi1}
	\Pi_{\mu\nu}(q^2) = \mi \int \dd^4 x \, \mathrm{e}^{\mi q \cdot x} 
	\bra{\Omega}\mathbb{T}\{j_{\mu}(x), j_{\nu}^{\dagger}(0)\}\ket{\Omega},
\end{equation}
where $j(x)$ and $j_{\mu}(x)$ denote the interpolating currents corresponding to the hexaquark states with $J=0$ and $J=1$, respectively, and $\ket{\Omega}$ represents the physical QCD vacuum.  

The two-point correlation function of the tensor type $\Pi_{\mu\nu}(q^2)$ contains contributions from both spin-$0$ and spin-$1$ degrees of freedom, and can be decomposed as
\begin{equation}
	\Pi_{\mu\nu}(q^2) = 
	-\left(g_{\mu\nu} - \frac{q_{\mu}q_{\nu}}{q^2}\right)\Pi_1(q^2) 
	+ \frac{q_{\mu}q_{\nu}}{q^2}\Pi_0(q^2),
\end{equation}
where the subscripts $1$ and $0$ correspond to spin-$1$ and spin-$0$ states, respectively.  By applying a projection, the spin-$0$ contribution can be removed, obtaining
\begin{equation}
	\Pi_1(q^2) = -\frac{1}{3}\left(g^{\mu\nu} - \frac{q^{\mu}q^{\nu}}{q^2}\right)\Pi_{\mu\nu}(q^2),
\end{equation}
which corresponds to the two-point correlation function of the $J=1$ hexaquark states.

\subsubsection{OPE Side}

In our calculation, the limit $m_u = m_d \to 0$ is taken such that isospin symmetry is preserved and no distinction is made between the two light flavors $q=u,d$.  The full QCD propagator, which incorporates both perturbative and non-perturbative contributions at all orders of vacuum condensates, is regarded. The quantities $\mathcal{S}_{q}^{jk}$ denote the full propagators of the $u,d$ quarks, whose explicit expressions are given as follows:
\begin{equation}
	\begin{aligned}
		\mathrm{i}\mathcal{S}_{q}^{jk}(x) =& \mathrm{i} \delta^{jk} \frac{\dslah{x}}{2\pi^2 x^4} 
		- \delta^{jk} m_q \frac{1}{4\pi^2 x^2} 
		- \mathrm{i} t^{jk}_a \frac{G^a_{\alpha\beta}}{32 \pi^2 x^2} 
		\left( \sigma^{\alpha\beta} \dslah{x} + \dslah{x} \sigma^{\alpha\beta} \right) 
		- \delta^{jk} \frac{\langle \bar{q} q \rangle}{12} 
		+ \mathrm{i} \delta^{jk} \frac{\dslah{x}}{48} m_q \langle \bar{q} q \rangle 
		\\
		&- \delta^{jk} \frac{x^2}{192} \langle g_s \bar{q} \sigma \cdot G q \rangle 
		+ \mathrm{i} \delta^{jk} \frac{x^2 \dslah{x}}{1152} m_q \langle g_s \bar{q} \sigma \cdot G q \rangle 
		- t^{jk}_a \frac{\sigma_{\alpha\beta}}{192} \langle g_s \bar{q} \sigma \cdot G q \rangle\\ 
		&- \mathrm{i} t^{jk}_a \frac{1}{768} 
		\left( \sigma_{\alpha\beta} \dslah{x} + \dslah{x} \sigma_{\alpha\beta} \right) 
		m_q \langle g_s \bar{q} \sigma \cdot G q \rangle.
	\end{aligned}
	\label{full-prop}
\end{equation}
For heavy quarks, only gluon condensates need to be considered, the full propagator of heavy quarks $\mathcal{S}_{Q}^{jk}$ is expressed in momentum space
\begin{equation}
	\begin{aligned}
		\mathcal{S}^{j k}_Q(p) & =\frac{\mi \delta^{j k}\left(\dslah p+m_Q\right)}{p^2-m_Q^2}-\frac{\mi}{4} \frac{t^{j k}_a G_{\alpha \beta}^a}{\left(p^2-m_Q^2\right)^2}\left[\sigma^{\alpha \beta}\left(\dslah p+m_Q\right)+\left(\dslah p+m_Q\right) \sigma^{\alpha \beta}\right] \\
		& +\frac{\mi \delta^{j k} m_Q\left\langle g_s^2 G^2\right\rangle}{12\left(p^2-m_Q^2\right)^3}\left[1+\frac{m_Q\left(\dslah p+m_Q\right)}{p^2-m_Q^2}\right] \\
		& +\frac{\mi \delta^{j k}}{48}\left\{\frac{\left(\not p+m_Q\right)\left[\dslah p\left(p^2-3 m_Q^2\right)+2 m_Q\left(2 p^2-m_Q^2\right)\right]\left(\dslah p+m_Q\right)}{\left(p^2-m_Q^2\right)^6}\right\}\left\langle g_s^3 G^3\right\rangle.
	\end{aligned}
\end{equation}
Here, $Q=c,b$ and $j,k$ denote the color indices. More details on the full propagator can be found in Refs.~\cite{Wang:2013vex,Albuquerque:2012jbz}.

Using the full propagators, one can analytically evaluate the correlation functions given in Eqs.~\eqref{eq:Pi0}--\eqref{eq:Pi1} by Wick's theorem. The correlation functions can be expressed as
\begin{equation}
	\begin{aligned}
		\Pi_{(\mu\nu)}(q^2)=&-\mi\varepsilon_{abc}\varepsilon_{a_1b_1c_1}\varepsilon_{def}\varepsilon_{d_1e_1f_1}\int_X\int_{P}\Tr\left[\mathcal{S}^{aa_1}_{d}(-x)\Gamma_1\Gamma_{(\mu)}\Gamma_1\mathcal{S}^{f_1f}_{Q}(p_1)\Gamma_1\Gamma_{(\nu)}\Gamma_1\right]\times\\
		&\Tr\left[\mathcal{C}\mathcal{S}^{Tcc_1}_{u}(-x)\mathcal{C}\Gamma_2\mathcal{S}^{bb_1}_{Q}(-p_2)\Gamma_2\right]\times\Tr\left[\mathcal{C}\mathcal{S}^{Td_1d}_{u}(x)\mathcal{C}\Gamma_2\mathcal{S}^{e_1e}_{d}(x)\Gamma_2\right].
	\end{aligned}
\end{equation}
For $\gamma$-matrices notation,  $\Gamma_1=\gamma_5,\Gamma_2=\mathbbm{1}$ denote the Type-I baryonic current Eq.~\eqref{eq:eta1}, while $\Gamma_1=\mathbbm{1},\Gamma_2=\gamma_5$ denote the Type-II baryonic current Eq.~\eqref{eq:eta2}. $\Gamma_{(\mu)}=\mi\gamma_5,\mathbbm{1},\gamma_{\mu},\gamma_{\mu}\gamma_{5}$ are corresponding to quantum numbers $J^{P} = 0^{-}, 0^{+}, 1^{-}, 1^{+}$, respectively. The simplified integration measure
\begin{equation}
	\int_X\int_{P}=\int\dd^4 x \int\frac{\dd^4 p_1}{(2\pi)^4} \int\frac{\dd^4 p_2}{(2\pi)^4}
\end{equation}
is also defined.

Through the Källén-Lehmann spectral representation
\begin{equation}
	\rho(s) = \frac{1}{\pi} \Im \Pi(s),
\end{equation}
one can correspond the correlation functions given in Eqs.~\eqref{eq:Pi0}-\eqref{eq:Pi1} to the spectral density and derive the spectral density in the form of the operator product expansion (OPE), which separates and factorizes the contribution from short distance (Wilson coefficients) and long distance (vacuum condensates). The spectral density in this work are retained up to dimension-12 operators, which can generally be expressed as
\begin{equation}\label{eq:OPE}
	\begin{aligned}
		\rho^{\mathrm{OPE}}(s) = &\  \rho^{\mathrm{pert}}(s) 
		+ \rho^{\langle \bar{q} q \rangle}(s) 
		+ \rho^{\langle G^2 \rangle}(s) 
		+ \rho^{\langle \bar{q} G q \rangle}(s) 
		+ \rho^{\langle \bar{q} q \rangle^2}(s) 
		+ \rho^{\langle G^3 \rangle}(s)+\rho^{\langle \bar{q} q \rangle \langle G^2 \rangle}(s)  \\
		&+ \rho^{\langle \bar{q} q \rangle \langle\bar{q}Gq\rangle}(s)+ \rho^{\langle \bar{q} q \rangle^3}(s) + \rho^{\langle \bar{q} G q \rangle \langle G^2 \rangle}(s) 
		+ \rho^{\langle \bar{q} q \rangle^2 \langle G^2 \rangle}(s)
		+ \rho^{\langle \bar{q} G q \rangle^2}(s)\\
		&+ \rho^{\langle \bar{q} q \rangle^2 \langle\bar{q}Gq\rangle}(s)
		+ \rho^{\langle \bar{q} q \rangle^4}(s).
	\end{aligned}
\end{equation}
Subsequently, through the dispersion relation, the spectral density on the OPE side can be used to express the correlation function $\Pi_{X,J^{P}}^{\text{OPE}}(q^2)$ as 
\begin{equation}\label{OPE}
	\Pi_{X,J^{P}}^{\text{OPE}}(q^2) = \int_{s_{\text{min}}}^{\infty} \dd s \ \frac{\rho_{X,J^{P}}^{\text{OPE}}(s)}{s - q^2},
\end{equation}
where $X$ denotes the corresponding ground hadronic state and $J^{P}$ denotes its quantum number; $s_{\text{min}}$ represents the kinematic threshold, typically corresponding to the sum of the masses of all quarks involved in the hadronic interpolating current, i.e. $s_{\text{min}}=4m_Q^2$ for these double-heavy hexaquark states. The analytical results of $\rho_{X,J^{P}}^{\text{OPE}}(s)$ are shown in the appendices. In the practical calculation of $\rho_{X,J^{P}}^{\text{OPE}}(s)$, the loop integrals of the relevant Feynman diagrams can be evaluated by the Schwinger parametrization method, and the ultraviolet divergences arising from these loop integrals are removed through renormalization in the $\overline{\text{MS}}$ scheme \cite{Albuquerque:2012jbz}. The relevant Feynman diagrams can be referred to Refs.~\cite{Wan:2019ake,Zhang:2024ulk}.

\subsubsection{Phenomenological Side}
In the phenomenological framework, the contributions from the ground state and the excited states (including the continuum spectrum) can be separated as  
\begin{equation}
	\rho_{X,J^{P}}^{\text{Phen}}(s) = \lambda_{X,J^{P}}^2 \, \delta\!\left(s - M_{X,J^{P}}^2\right) 
	+ \theta\!\left(s - s_0\right) \rho_{X,J^{P}}(s),
\end{equation}
where $m_{X,J^{P}}$ denotes the mass of the ground state, and $s_0$ is the threshold parameter, which characterizes the onset of the excited states and the continuum spectrum.  The decay constants $\lambda_{X,J^{P}}$ of the ground state are defined as
\begin{equation}
	\begin{aligned}
		\lambda_{X,0^{\pm}}&=\bra{\Omega}j(x)\ket{X},\\
		\lambda_{X,1^{\pm}}\varepsilon_{\mu}&=\bra{\Omega}j_{\mu}(x)\ket{X},
	\end{aligned}
\end{equation}
which reflect the coupling of the interpolating currents and the ground state hadronic states.

By applying the dispersion relation, the phenomenological representation of the correlation function can be written as  
\begin{equation}\label{phen}
	\Pi_{X,J^{P}}^{\text{Phen}}(q^2) 
	= \frac{\lambda_{X,J^{P}}^2}{M_{X,J^{P}}^2 - q^2} 
	+ \int_{s_0}^{\infty} \dd s \, \frac{\rho_{X,J^{P}}(s)}{s - q^2},
\end{equation}
where the first term corresponds to the pole contribution of the ground state, while the second term accounts for the contributions from the excited states and the continuum spectrum.

\subsection{Hadronic Mass and Decay Constant}
According to the hypothesis of quark-hadron duality, the correlation functions obtained from the OPE representation and the phenomenological representation should be consistent. In particular, the spectral densities from the two sides are expected to be approximately equal above the continuum threshold parameter $s_0$. Based on this assumption, we can combine Eqs.~\eqref{OPE} and \eqref{phen}. By performing a Borel transformation on both sides of the equation, the contributions from the excited states and the continuum spectrum are exponentially suppressed, leading to
\begin{equation}\label{SR}
	\lambda_{X,J^{P}}^2 \, \mathrm{e}^{-M_{X,J^{P}}^2 / M_B^2} 
	= \int_{s_{\min}}^{s_0} \dd s \, \rho_{X,J^{P}}^{\text{OPE}}(s) \, \mathrm{e}^{-s / M_B^2}.
\end{equation}

From the sum rule given in Eq.~\eqref{SR}, the mass of the ground-state hadron $X$ can be expressed as
\begin{equation}\label{mass}
	M_{X,J^{P}}(s_0, M_B^2) 
	= \sqrt{-\frac{L_{X,J^{P},1}(s_0, M_B^2)}{L_{X,J^{P},0}(s_0, M_B^2)}},
\end{equation}
where
\begin{equation}\label{moment}
	\begin{aligned}
		L_{X,J^{P},0}(s_0, M_B^2) 
		&= \int_{s_{\min}}^{s_0} \dd s \, \rho^{\text{OPE}}(s) \, \mathrm{e}^{-s / M_B^2} , \\
		L_{X,J^{P},1}(s_0, M_B^2) 
		&= \frac{\partial}{\partial (M_B^{-2})} L_{X,J^{P},0}(s_0, M_B^2).
	\end{aligned}
\end{equation}
Furthermore, the decay constant can be extracted from Eq.~\eqref{SR} as
\begin{equation}
	\lambda_{X,J^{P}}(s_0, M_B^2) 
	= \sqrt{\mathrm{e}^{M_{X,J^{P}}^2(s_0, M_B^2)/M_B^2} \, L_{X,J^{P},0}(s_0, M_B^2)}.
\end{equation}

\section{Numerical Analysis\label{na}}

\subsection{Input Parameters}
In the numerical calculations of QCDSR, the following input parameters are adopted \cite{ParticleDataGroup:2024cfk,Colangelo:2000dp,Tang:2019nwv,Wan:2019ake,Wan:2021vny}, where $q$ represents the $u,d$ quarks:
\begin{equation}
	\begin{array}{ll}
		\langle \bar{q}q \rangle = -(0.24 \pm 0.01)^3 \ \text{GeV}^3, &
		\langle g_s^2 G^2 \rangle = (0.88 \pm 0.25) \ \text{GeV}^4, \\ 
		\langle g_s^3 G^3 \rangle = (0.045 \pm 0.013) \ \text{GeV}^6, &
		\langle \bar{q}g_s \sigma \cdot G q \rangle = m_0^2 \langle \bar{q}q \rangle,\\
		\bar{m}_c(\bar{m}_c)=1.273\pm 0.0028 \mathrm{GeV}, & \bar{m}_b(\bar{m}_b)=4.183\pm 0.004 \mathrm{GeV},\\
	\end{array}
\end{equation}
Here, the $\overline{\mathrm{MS}}$ mass of $c,b$ quarks from the latest PDG results \cite{ParticleDataGroup:2024cfk} are used, and the value of ratio $\langle \bar{q}g_s \sigma \cdot G q \rangle / \langle \bar{q}q \rangle $ is $m_0^2 = (0.8 \pm 0.1) \ \text{GeV}^2$ .

\subsection{Mass Stability}

In establishing the theoretical framework of QCDSR, two additional parameters, $s_0$ and $M_B$, are introduced. However, all physical observables, such as the hadronic mass $M_X$, should not depend on these parameters. Therefore, it is necessary to identify suitable ranges of $s_0$ and $M_B$ such that the variation of $M_X$ with respect to $s_0$ and $M_B$ is minimized. The region in which the mass exhibits minimal sensitivity to these parameters is referred to as the Borel window. A reliable Borel window must also satisfy two additional conditions.

First, we require the OPE to be convergent, which means that within the Borel window the contribution from the higher-dimensional condensate should be as small as possible. We define the ratio of the dimension-$n$ condensate contribution as
\begin{equation}
	R^{\langle \mathcal{O}_{n}\rangle}_{X,J^{P}} =\left\vert \frac{L^{\langle \mathcal{O}_{n}\rangle}_{X,J^{P},0}(s_0,M_B^2)}{L_{X,J^{P},0}(s_0,M_B^2)}\right\vert,
\end{equation}
where
\begin{equation}
	L^{\langle \mathcal{O}_{n}\rangle}_{X,J^{P},0}(s_0,M_B^2)=\int_{s_{\min}}^{s_0} \dd s \ \rho^{\langle \mathcal{O}_{n}\rangle}(s) \mathrm{e}^{-s / M_B^2} .
\end{equation}
For conventional hadrons, the OPE typically converges relatively rapidly, and in many cases contributions up to dimension $n\leqslant 6$ are sufficient to demonstrate convergence \cite{Colangelo:2000dp}. For multiquark states, however, higher-dimensional operators are often needed to be taken into account. In the present analysis, the OPE is performed up to condensates of dimension 12, ensuring a satisfactory convergence of the series within the chosen Borel window. To ensure convergence, we require that the highest-dimensional contribution satisfies 
$R^{\langle \mathcal{O}_{12}\rangle}_{X,J^{P}} \lesssim 10\%$ 
within the working Borel window.  In addition, we have explicitly evaluated the representative dimension-13 contribution 
$\langle \bar{q} q \rangle \langle \bar{q} G q \rangle^2$, 
and find that its numerical effect is negligible in the relevant Borel region.  This confirms that the truncation of the OPE at dimension 12 is self-consistent and that the series exhibits satisfactory convergence.

Second, since the quantity being extracted is the mass of the ground-state hadron, the pole contribution associated with the ground state should dominate the spectral density, while the contributions from excited states and the continuum should remain suppressed. We define the ratio of the pole contribution as
\begin{equation}
	R^\text{PC}_{X,J^{P}} = \frac{L_{X,J^{P},0}(s_0,M_B^2)}{L_{X,J^{P},0}(\infty,M_B^2)}.
\end{equation}

In the QCDSR framework, the pole contribution is intrinsically tied to the quark--hadron duality ansatz, where the spectral density above $s_0$ is approximated by its perturbative expression. For conventional hadrons, the spectral structure is relatively simple, and the separation between the lowest-lying pole and the continuum is fairly distinct. This typically leads to a pole contribution of about $50\%$, ensuring a clear ground-state dominance.

For multiquark systems, however, the situation becomes qualitatively different. The interpolating currents can couple not only to the genuine multiquark configurations but may also to two-hadron scattering states and other higher Fock components of the hadrons with the same quantum numbers. As a consequence, the spectral strength is more broadly distributed and extends further into the higher-energy region, which reflect on the enhancement of the relative weight of large-$s$ contributions in the dispersion relation and naturally reduces the pole fraction \cite{Chen:2014vha}.

Accordingly, QCDSR analyses of multiquark states generally adopt somewhat lower pole contributions, provided that Borel stability and OPE convergence are simultaneously satisfied. This feature reflects the intrinsic structure of the spectral density rather than a limitation of the method itself. Typical choices in the literature include $R^\text{PC}_{X,J^{P}}>40\%$ \cite{Wang:2015epa} or even $R^\text{PC}_{X,J^{P}}>20\%$ \cite{Agaev:2022ast}. For the present hexaquark system, the effect is further amplified due to the higher dimensionality of the correlation function. By requiring a consistent Borel window with good convergence and stability, we find that $R^\text{PC}_{X,J^{P}}>30\%$ provides a self-consistent working criterion. More restrictive conditions do not lead to a stable sum rule in this case.

\subsection{Determine the Borel Window}

To determine the explicit Borel window, we follow the method described in Ref.~\cite{Albuquerque:2012jbz}. For the $\Lambda_Q\bar{\Sigma}_Q$ states, the extracted mass $M_X$ and the pole contribution $R^{\mathrm{PC}}$ are examined over sufficiently broad ranges of $\sqrt{s_0}$ and $M_B^2$. According to the latest PDG compilation \cite{ParticleDataGroup:2024cfk}, the baryon–antibaryon thresholds are approximately $E_{\mathrm{th},c}=4.71~\mathrm{GeV}$ for $\Lambda_c \bar{\Sigma}_c$ and $E_{\mathrm{th},b}=11.43~\mathrm{GeV}$ for $\Lambda_b \bar{\Sigma}_b$. The continuum threshold parameter $\sqrt{s_0}$ is therefore scanned starting from the corresponding physical threshold. For the charm sector, we consider $4.7~\mathrm{GeV}\leqslant \sqrt{s_0} \leqslant 7~\mathrm{GeV}$ and $3~\mathrm{GeV}^2 \leqslant M_B^2 \leqslant 6~\mathrm{GeV}^2$, while for the bottom sector we explore $11.4~\mathrm{GeV}\leqslant \sqrt{s_0} \leqslant 13~\mathrm{GeV}$ and $6.5~\mathrm{GeV}^2 \leqslant M_B^2 \leqslant 10.5~\mathrm{GeV}^2$. It is found that the numerical behaviors of the Type-I currents are mutually similar and clearly distinct from those of the Type-II currents, and representative examples are therefore displayed for one typical Type-I state and one typical Type-II state.

It is important to emphasize that, within the QCDSR framework, the continuum threshold $\sqrt{s_0}$ is an effective parameter introduced under the assumption of quark–hadron duality to approximate the onset of continuum contributions. It should not be identified with the physical threshold of specific hadronic scattering states, nor with the simple sum of constituent hadron masses.

For conventional hadrons, the spectral structure is relatively clean, and the effective threshold is often close to the squared mass of the first excited state. In such cases, the transition from the ground-state pole dominance to the continuum is comparatively sharp. In contrast, for multiquark systems, the spectral distribution is typically more diffuse. As a result, the separation between the pole and continuum contributions becomes less distinct.

In practical QCDSR analyses, $\sqrt{s_0}$ is therefore taken to be slightly above the extracted ground-state mass $M_X$, and is commonly parameterized as $\sqrt{s_0} = M_X + \delta$. For conventional hadrons, one typically has $\delta \approx 0.5~\mathrm{GeV}$ \cite{Colangelo:2000dp}, while for multiquark systems a somewhat broader range is expected. In the present baryonium systems, we adopt $0.5 \leqslant \delta \leqslant 0.9~\mathrm{GeV}$. These values emerge from the combined requirements of Borel stability and pole dominance rather than being imposed a priori.

\begin{figure}[H]
	\centering
	\subfigure[]{\includegraphics[width=0.45\textwidth]{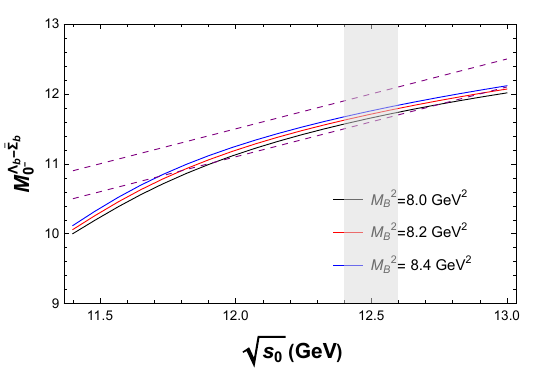}}
	\subfigure[]{\includegraphics[width=0.45\textwidth]{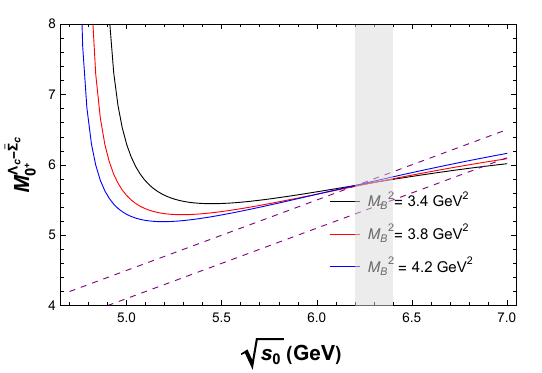}}
	\caption{
		Dependence of the extracted mass $M_X$ on $\sqrt{s_0}$ for the $0^-$ Type-I $\Lambda_b\bar{\Sigma}_b$ state (panel a) and the $0^+$ Type-II $\Lambda_c\bar{\Sigma}_c$ state (panel b); in each panel, the three curves correspond to representative $M_B^2$ values within the allowed Borel window, the dashed lines indicate $\sqrt{s_0}=M_X+0.5~\mathrm{GeV}$ and $\sqrt{s_0}=M_X+0.9~\mathrm{GeV}$, and the shaded region highlights the optimal $\sqrt{s_0}$ range determined from Borel stability and pole dominance.
	}
	\label{fig:Ms0}
\end{figure}

Fig.~\ref{fig:Ms0} shows the corresponding $\sqrt{s_0}$ dependence of $M_X$ for the $0^-$ Type-I $\Lambda_b\bar{\Sigma}_b$ state and the $0^+$ Type-II $\Lambda_c\bar{\Sigma}_c$ state at fixed $M_B^2$. For the Type-I $0^-$ $\Lambda_b\bar{\Sigma}_b$ state, a value of $\sqrt{s_0}$ close to threshold leads to a larger sensitivity $\partial M_X / \partial \sqrt{s_0}$. At $\sqrt{s_0}\approx 12.5~\mathrm{GeV}$, a flatter slope together with a moderate mass gap $\delta=\sqrt{s_0}-M_X$ occurs. A similar analysis for the Type-II $0^+$ $\Lambda_c\bar{\Sigma}_c$ state shows that values of $\sqrt{s_0}$ very close to threshold produce unstable behavior, with almost vertical drop features indicating a near-singular sensitivity of $\partial M_X / \partial \sqrt{s_0}$. When $\sqrt{s_0}\approx 6.3~\mathrm{GeV}$, the mass exhibits the weakest $M_B^2$ dependence and yields a reasonable mass gap $\delta=\sqrt{s_0}-M_X$. These considerations suggest the optimal $\sqrt{s_0}$ ranges for the two representative cases. Values of $\sqrt{s_0}$ very close to the baryon--antibaryon threshold lead to a rapid variation of $M_X$, which is likely caused by the increasing influence of scattering states in the continuum. Such behaviour prevents the formation of a stable Borel platform.

Taking OPE convergence and pole dominance into account, the admissible Borel window is then determined self-consistently. After selecting the proper $\sqrt{s_0}$, within a mass plateau exhibiting $M_B^2$ stability, the lower bound of $M_B^2$ is set by requiring satisfactory convergence of the OPE series, namely that the dimension-12 contribution remains sufficiently small ($R^{\langle \mathcal{O}_{12}\rangle}_{X,J^{P}} \lesssim 10\%$) while the dimension-13 term does not spoil the hierarchical structure. The upper bound of $M_B^2$ is constrained by pole dominance through $R^{\mathrm{PC}}_{X,J^{P}}>30\%$, and the final working interval of $M_B^2$ is obtained from the simultaneous fulfillment of these criteria. In this work, $\sqrt{s_0}$ is allowed to vary within $\pm0.1~\mathrm{GeV}$ around the selected central value, and the resulting Borel window satisfies $\Delta M_B^2 > 0.5~\mathrm{GeV}^2$ to ensure numerical stability.

\subsection{Numerical Results}

\begin{figure}[H]
	\centering
	\subfigure[]{\includegraphics[width=0.45\textwidth]{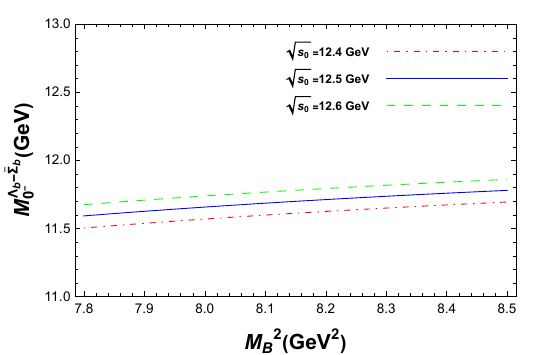}}
	\subfigure[]{\includegraphics[width=0.45\textwidth]{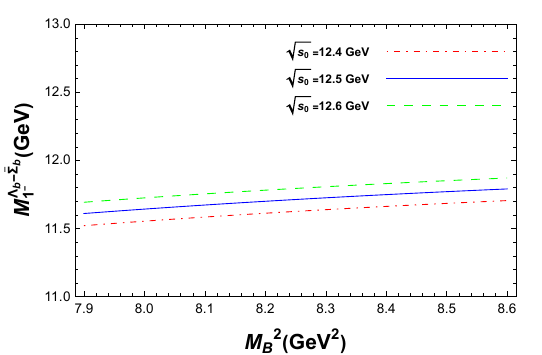}}
	\subfigure[]{\includegraphics[width=0.45\textwidth]{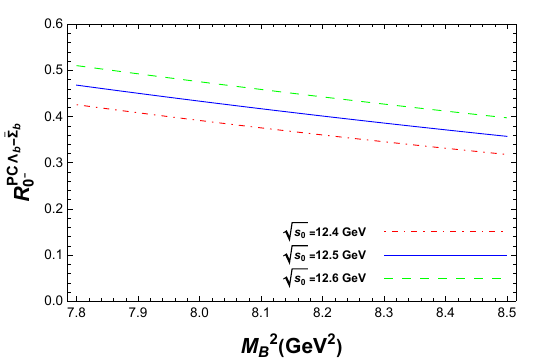}}
	\subfigure[]{\includegraphics[width=0.45\textwidth]{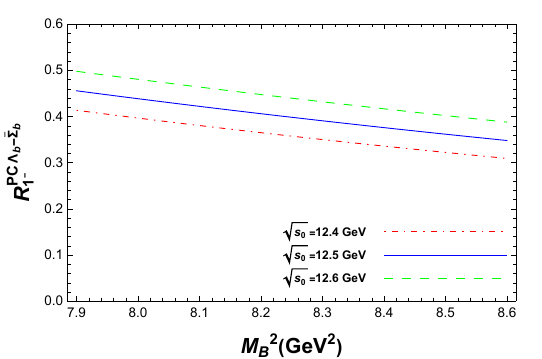}}
	\subfigure[]{\includegraphics[width=0.45\textwidth]{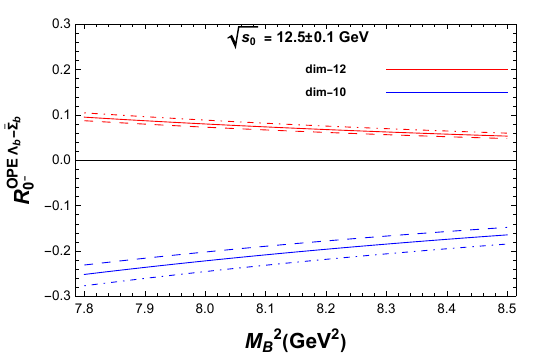}}
	\subfigure[]{\includegraphics[width=0.45\textwidth]{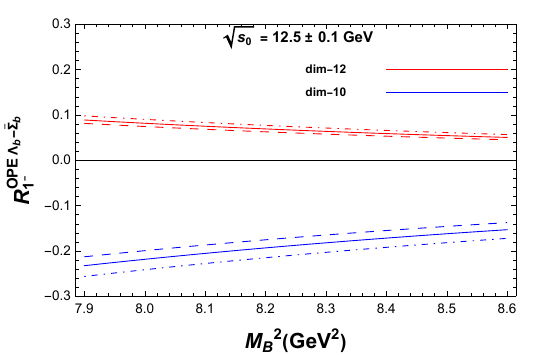}}
	
	\caption{The figures for $0^-$ and $1^-$ states coupled to Type-I currents}
	\label{fig:10-}
\end{figure}

Finally, totally 10 baryonium states are found. For the baryonium currents constructed by the Type-I baryonic interpolating currents, we find that the states with $J^P = 0^-$ and $1^-$ admit reliable Borel windows for the bottom sector; but for the charm sector, no Borel windows can be obtained unless broaden the pole dominant condition to $R^{\text{PC}}_{X,J^{P}} > 15\%$, which would be unreliable. For the states with $J^P = 0^+$ and $1^+$, no reliable Borel window can be identified, suggesting that such configurations do not couple effectively to the currents employed in this work. The corresponding numerical results are summarized in Table.~\ref{1mass}. Fig.~\ref{fig:10-} presents the relevant curves for the hidden-bottom baryonium states. Panels (a) and (b) shows the masses for the $0^-$ and $1^-$ states, respectively; panels.~(c) and (d) displays the pole contribution $R^{\text{PC}}$ for the $0^-$ and $1^-$ states, respectively; panels.~(e) and (f) illustrate the ratios $R^{\langle \mathcal{O}_n \rangle}$ for $n = 10, 11, 12$ in the $0^-$ and $1^-$ states. Since the contribution from the dimension 11 condensate is identically zero in this channel, it is not shown in the figure. 

\begin{table}[H]
	\centering
		\linespread{1.05}\selectfont
		\begin{tabular}{cccccccc}
			\hline\hline
			$J^{P}$ &  State & $\sqrt{s_0}(\mathrm{GeV})$ & $M_B^2 (\mathrm{GeV}^2)$ & $M_X(\mathrm{GeV})$ & $\lambda_{X}(\mathrm{GeV}^8)$ & $R^{\mathrm{PC}}(\%)$ & $\abs{R^{\langle \mathcal{O}_{12}\rangle}}(\%)$\\
			\hline
			$0^{-}$ 
			&$\Lambda_b\bar{\Sigma}_b$&$12.5\pm 0.1$ & $7.8-8.5$ & $11.68\pm 0.18$& $(5.8\pm 1.7)\times 10^{-3}$& $32-51$ & $4.8-10.4$ \\
			\hline
			$1^{-}$ 
			& $\Lambda_b\bar{\Sigma}_b$ & $12.5\pm 0.1$ & $7.9-8.6$ & $11.70\pm 0.18$&$(5.8\pm 1.7)\times 10^{-3}$ & $31-50$&$4.5-10.0$ \\
			\hline\hline
		\end{tabular}
	\caption{The related numerical results of the Type-I currents}
	\label{1mass}
\end{table}
\begin{table}[h]
	\centering
		\linespread{1.05}\selectfont
		\begin{tabular}{cccccccc}
			\hline\hline
			$J^{P}$ &  State & $\sqrt{s_0}(\mathrm{GeV})$ & $M_B^2 (\mathrm{GeV}^2)$ & $M_X(\mathrm{GeV})$ & $\lambda_{X}(\mathrm{GeV}^8)$ & $R^{\mathrm{PC}}(\%)$ & $\abs{R^{\langle \mathcal{O}_{12}\rangle}}(\%)$\\
			\hline
			$0^{-}$ & $\Lambda_c\bar{\Sigma}_c$ & $6.3\pm 0.1$ & $4.1-4.7$ & $5.72\pm 0.08$ & $(1.7\pm 0.3)\times 10^{-3}$& $31-54$&$0.5-1.2$ \\
			&$\Lambda_b\bar{\Sigma}_b$&$12.6\pm 0.1$ & $9.1-10.5$ & $11.86\pm 0.10$& $(1.4\pm 0.2)\times 10^{-2}$& $30-54$ & $0.4-1.0$ \\
			\hline
			$0^{+}$ & $\Lambda_c\bar{\Sigma}_c$ & $6.3\pm 0.1$ & $3.3-4.3$ & $5.77\pm 0.06$ & $(1.4\pm 0.1)\times 10^{-3}$& $32-69$&$1.5-8.3$ \\
			&$\Lambda_b\bar{\Sigma}_b$&$12.5\pm 0.1$ & $7.5-9.3$ & $11.89\pm 0.07$& $(9.2\pm 1.3)\times 10^{-3}$& $30-63$ & $1.7-9.5$ \\
			\hline
			$1^{-}$ & $\Lambda_c\bar{\Sigma}_c$  & $6.4\pm 0.1$ & $4.2-4.9$& $5.79\pm 0.09$  & $(2.0\pm 0.3)\times 10^{-3}$ & $31-56$&$0.4-1.0$ \\
			& $\Lambda_b\bar{\Sigma}_b$ & $12.7\pm 0.1$ & $9.1-10.6$ & $11.92\pm 0.11$&$(1.5\pm 0.2)\times 10^{-2}$ & $33-57$&$0.3-1.0$ \\
			\hline
			$1^{+}$ & $\Lambda_c\bar{\Sigma}_c$  & $6.4\pm 0.1$ & $3.3-4.5$& $5.82\pm 0.09$  & $(1.4\pm 0.3)\times 10^{-3}$ & $31-73$&$1.2-8.6$ \\
			& $\Lambda_b\bar{\Sigma}_b$ & $12.6\pm 0.1$ & $7.6-9.5$ & $11.95\pm 0.07$&$(1.0\pm 0.4)\times 10^{-2}$ & $31-65$&$1.5-8.5$ \\
			\hline\hline
		\end{tabular}
	\caption{The related numerical results of the Type-II currents}
	\label{2mass}
\end{table}

\begin{figure}[H]
	\centering
	\subfigure[]{\includegraphics[width=0.45\textwidth]{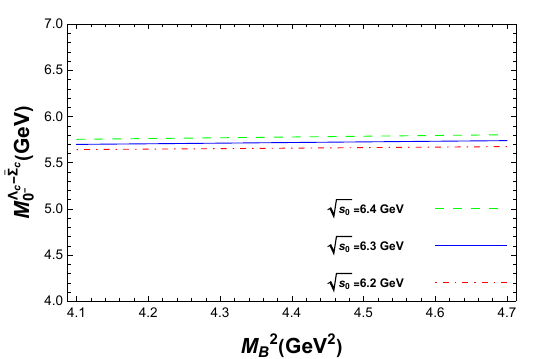}}
	\subfigure[]{\includegraphics[width=0.45\textwidth]{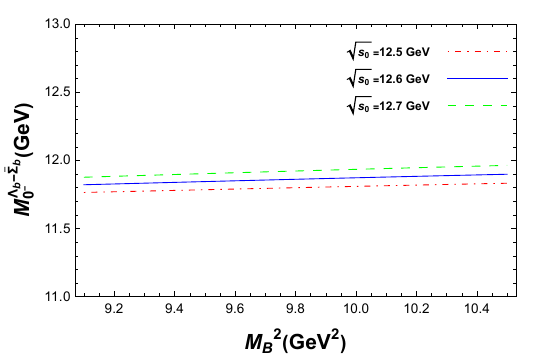}}
	\subfigure[]{\includegraphics[width=0.45\textwidth]{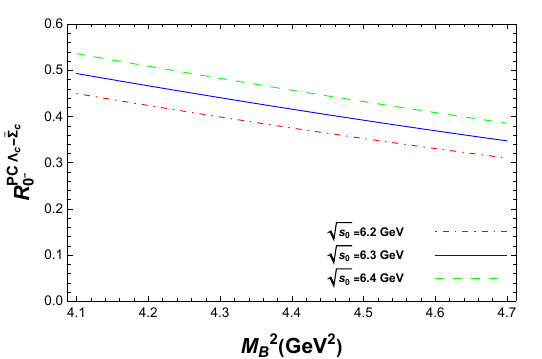}}
	\subfigure[]{\includegraphics[width=0.45\textwidth]{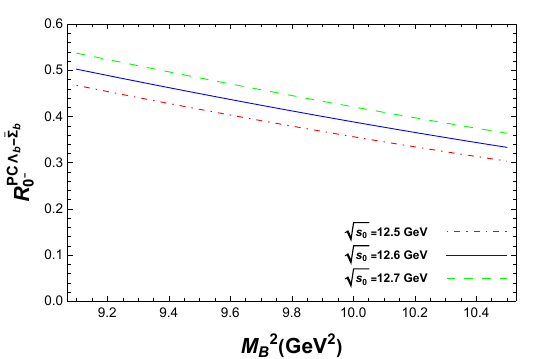}}
	\subfigure[]{\includegraphics[width=0.45\textwidth]{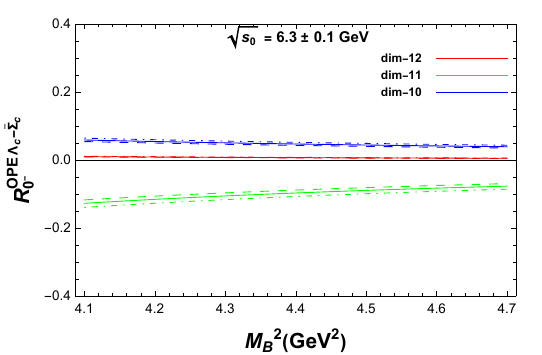}}
	\subfigure[]{\includegraphics[width=0.45\textwidth]{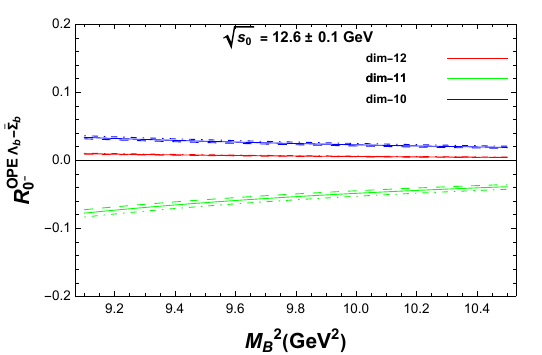}}
	
	\caption{The figures for $0^-$ states coupled to Type-II currents}
	\label{fig:20-}
\end{figure}

\begin{figure}[H]
	\centering
	\subfigure[]{\includegraphics[width=0.45\textwidth]{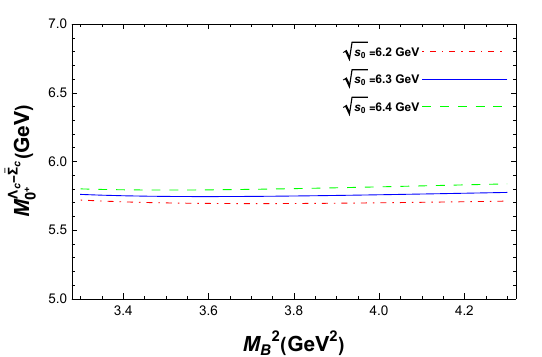}}
	\subfigure[]{\includegraphics[width=0.45\textwidth]{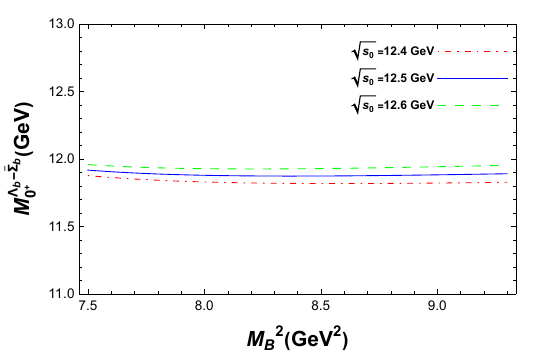}}
	\subfigure[]{\includegraphics[width=0.45\textwidth]{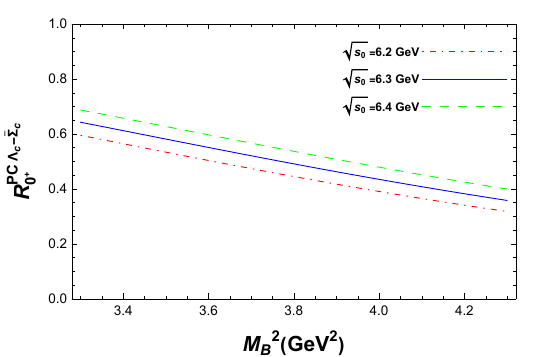}}
	\subfigure[]{\includegraphics[width=0.45\textwidth]{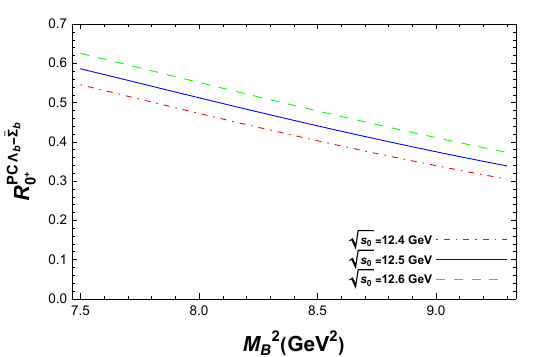}}
	\subfigure[]{\includegraphics[width=0.45\textwidth]{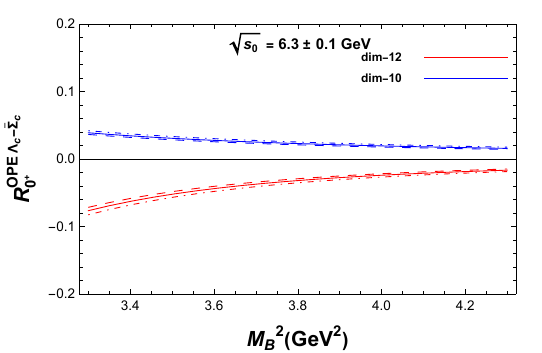}}
	\subfigure[]{\includegraphics[width=0.45\textwidth]{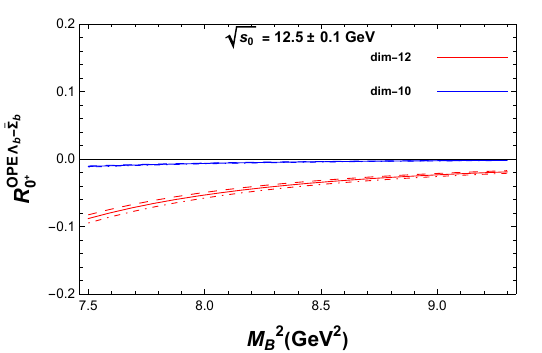}}
	
	\caption{The figures for $0^+$ states coupled to Type-II currents}
	\label{fig:20+}
\end{figure}

\begin{figure}[H]
	\centering
	\subfigure[]{\includegraphics[width=0.45\textwidth]{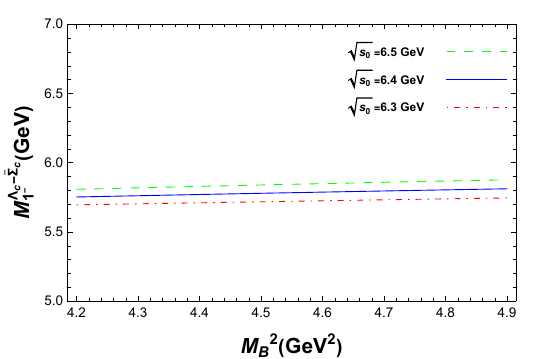}}
	\subfigure[]{\includegraphics[width=0.45\textwidth]{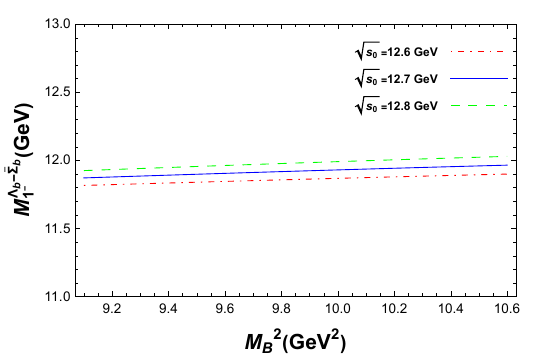}}
	\subfigure[]{\includegraphics[width=0.45\textwidth]{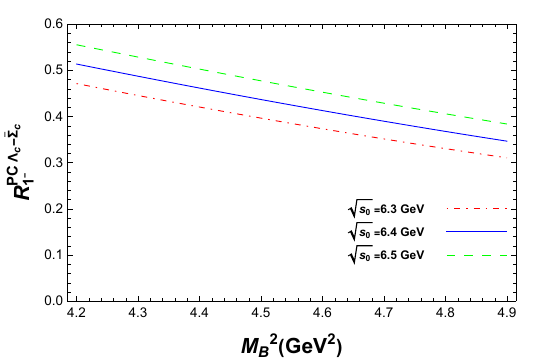}}
	\subfigure[]{\includegraphics[width=0.45\textwidth]{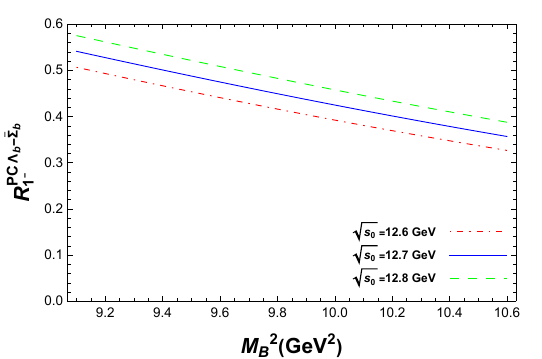}}
	\subfigure[]{\includegraphics[width=0.45\textwidth]{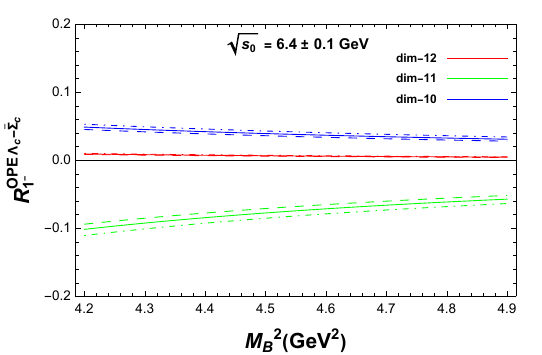}}
	\subfigure[]{\includegraphics[width=0.45\textwidth]{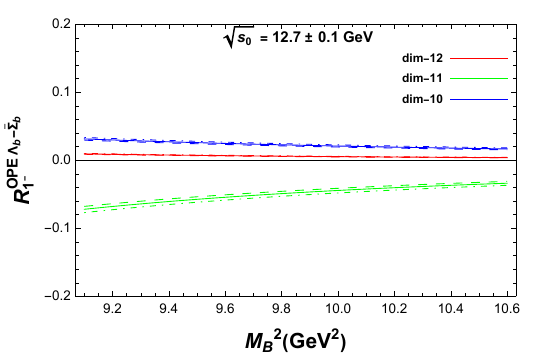}}
	
	\caption{The figures for $1^-$ states coupled to Type-II currents}
	\label{fig:21-}
\end{figure}

\begin{figure}[H]
	\centering
	\subfigure[]{\includegraphics[width=0.45\textwidth]{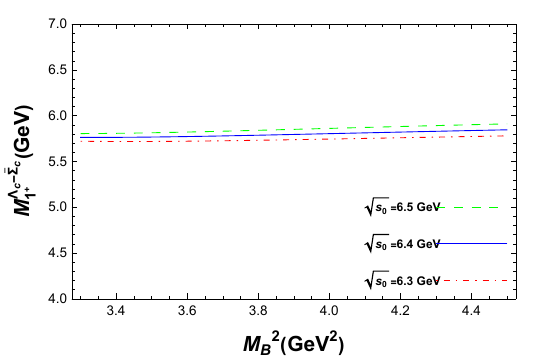}}
	\subfigure[]{\includegraphics[width=0.45\textwidth]{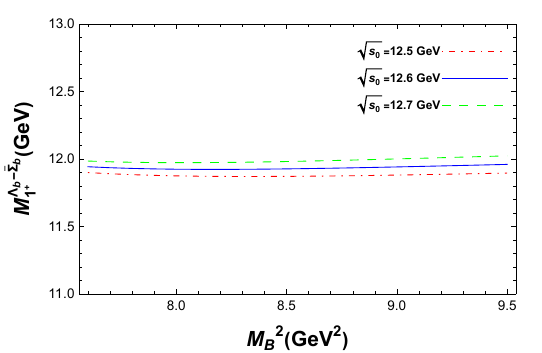}}
	\subfigure[]{\includegraphics[width=0.45\textwidth]{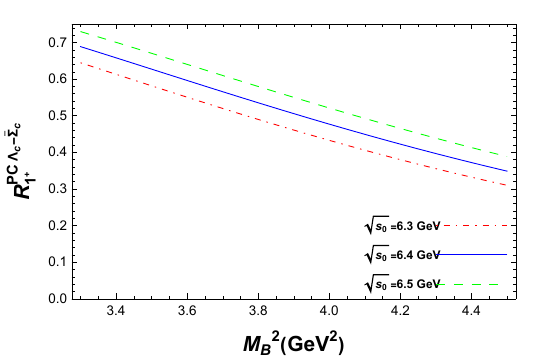}}
	\subfigure[]{\includegraphics[width=0.45\textwidth]{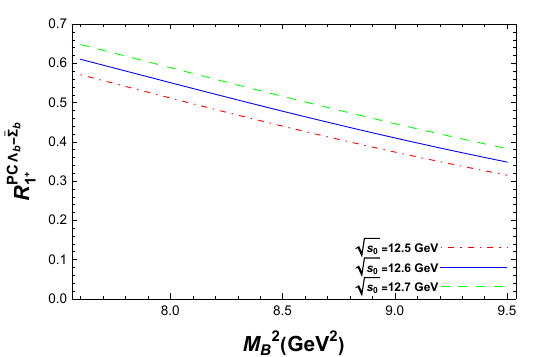}}
	\subfigure[]{\includegraphics[width=0.45\textwidth]{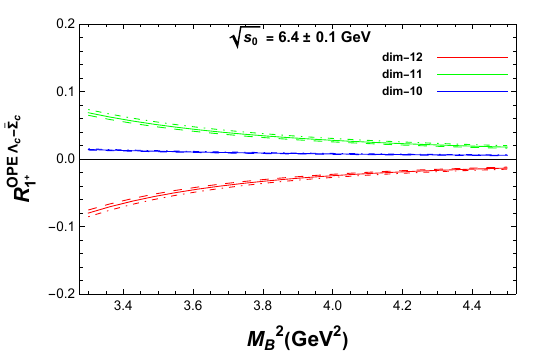}}
	\subfigure[]{\includegraphics[width=0.45\textwidth]{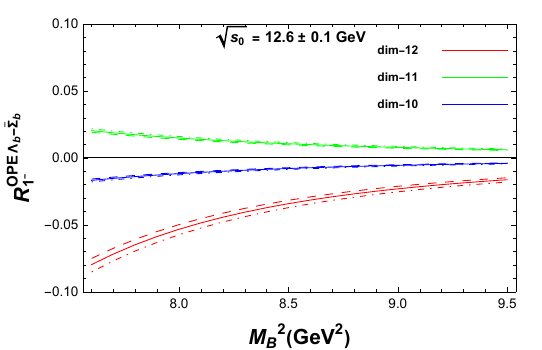}}
	
	\caption{The figures for $1^+$ states coupled to Type-II currents}
	\label{fig:21+}
\end{figure}

For the baryonium currents constructed by the Type-II baryonic interpolating currents, we find that the states with $J^P = 0^-, 0^+, 1^-, 1^+$ all exhibit relatively large and reliable Borel windows.  It can be found that for each of the $0^-$ and $1^-$ hidden-bottom states, the decay constants are larger than those in the Type-I case, indicating that the Type-II currents couple to the corresponding baryonium states better, corresponding to larger and flatter Borel windows. The corresponding numerical results are presented in Table.~\ref{2mass}. Figs.~\ref{fig:20-}--\ref{fig:21+} show the relevant curves for the $J^P = 0^-, 0^+, 1^-, 1^+$ states, where each sub-figure has the same meaning as in Fig.~\ref{fig:10-}. In the sub-figures displaying $\abs{R^{\langle \mathcal{O}_n \rangle}}$ for $n = 10, 11, 12$, the contribution from the dimension-11 condensate is omitted whenever it is identically zero.

As can be seen from Tables.~\ref{1mass},\ref{2mass}, both the extracted mass $M_X$ and the coupling $\lambda_X$ carry certain uncertainties, which originate from the variations in the parameters $s_0$ and $M_B$. Typically, the uncertainty in $M_X$ is of the order of $\Lambda_{\text{QCD}} \sim 200~\mathrm{MeV}$, which is consistent with our results. For multiquark systems, however, the uncertainties are generally expected to be larger than those of conventional hadrons due to the more involved spectral structure and the stronger sensitivity to the continuum threshold. In this sense, the QCDSR approach to multiquark configurations should be regarded as a semi-quantitative framework, aiming to capture the dominant features of the ground-state signal rather than providing high-precision predictions. A flatter Borel window leads to smaller uncertainties and indicates a stronger coupling between the chosen interpolating current and the corresponding ground hadronic state.

\section{Decay Modes Analyses \label{dm}}
In experimental studies, determining the internal structure of a hadronic state requires not only to observe its mass and quantum numbers, but also reconstructing the state from its decay products. Moreover, in realistic experimental environments, the observed $XYZ$ resonances are often not pure states. Instead, they are mixtures of several hadronic configurations with similar masses and identical quantum numbers. To distinguish among them, it is necessary to examine their possible decay channels. Therefore, it is important to analyze the potential decay modes of the hexaquark states of the type $\Lambda_Q \bar{\Sigma}_Q$, so as to provide useful guidance for their experimental identification.

\begin{figure}[h]

	\tikzset{every picture/.style={line width=0.75pt}} 
	
	\begin{tikzpicture}[x=0.75pt,y=0.75pt,yscale=-1,xscale=1]
		
		\draw  [fill={rgb, 255:red, 0; green, 0; blue, 0 }  ,fill opacity=0.69 ] (41.19,91.21) .. controls (55.12,91.19) and (66.44,118.59) .. (66.48,152.4) .. controls (66.51,186.21) and (55.25,213.63) .. (41.31,213.64) .. controls (27.38,213.66) and (16.06,186.26) .. (16.02,152.45) .. controls (15.99,118.64) and (27.25,91.22) .. (41.19,91.21) -- cycle ;
		\draw    (41.19,91.21) -- (141.81,90.31) ;
		\draw    (141.81,90.31) -- (216.21,213.56) ;
		\draw    (60.37,193.39) -- (141.3,193.39) ;
		\draw    (141.3,193.39) -- (216.21,213.56) ;
		\draw   (142.05,193.39) .. controls (141.59,192.05) and (141.3,190.32) .. (141.3,188.16) .. controls (141.3,177.39) and (148.51,177.39) .. (148.51,183.03) .. controls (148.51,188.67) and (141.3,188.67) .. (141.3,177.9) .. controls (141.3,167.13) and (148.51,167.13) .. (148.51,172.77) .. controls (148.51,178.41) and (141.3,178.41) .. (141.3,167.64) .. controls (141.3,156.88) and (148.51,156.88) .. (148.51,162.52) .. controls (148.51,168.16) and (141.3,168.16) .. (141.3,157.39) .. controls (141.3,146.62) and (148.51,146.62) .. (148.51,152.26) .. controls (148.51,157.9) and (141.3,157.9) .. (141.3,147.13) .. controls (141.3,136.36) and (148.51,136.36) .. (148.51,142) .. controls (148.51,147.64) and (141.3,147.64) .. (141.3,136.88) .. controls (141.3,126.11) and (148.51,126.11) .. (148.51,131.75) .. controls (148.51,137.39) and (141.3,137.39) .. (141.3,126.62) .. controls (141.3,115.85) and (148.51,115.85) .. (148.51,121.49) .. controls (148.51,127.13) and (141.3,127.13) .. (141.3,116.36) .. controls (141.3,105.59) and (148.51,105.59) .. (148.51,111.24) .. controls (148.51,116.88) and (141.3,116.88) .. (141.3,106.11) .. controls (141.3,95.34) and (148.51,95.34) .. (148.51,100.98) .. controls (148.51,106.62) and (141.3,106.62) .. (141.3,95.85) .. controls (141.3,93.52) and (141.64,91.69) .. (142.17,90.31) ;
		\draw  [fill={rgb, 255:red, 139; green, 87; blue, 42 }  ,fill opacity=1 ] (211.83,213.56) .. controls (211.83,211.15) and (213.79,209.2) .. (216.21,209.2) .. controls (218.63,209.2) and (220.59,211.15) .. (220.59,213.56) .. controls (220.59,215.96) and (218.63,217.92) .. (216.21,217.92) .. controls (213.79,217.92) and (211.83,215.96) .. (211.83,213.56) -- cycle ;
		\draw  [line width=1.5]  (90.18,88.45) -- (93.97,90.57) -- (90.18,92.69) ;
		\draw  [line width=1.5]  (166.63,126.84) -- (166.76,131.16) -- (162.98,129.02) ;
		\draw  [line width=1.5]  (192.37,208.98) -- (189.2,206.01) -- (193.4,204.87) ;
		\draw  [line width=1.5]  (103.77,195.26) -- (99.97,193.14) -- (103.77,191.01) ;
		\draw    (55.21,101.08) -- (173.08,99.54) ;
		\draw    (173.08,99.54) -- (218.1,153.39) ;
		\draw    (55.73,203.13) -- (172.74,203.64) ;
		\draw    (172.74,203.64) -- (218.1,153.39) ;
		\draw  [fill={rgb, 255:red, 74; green, 144; blue, 226 }  ,fill opacity=1 ] (213.72,153.39) .. controls (213.72,150.98) and (215.68,149.03) .. (218.1,149.03) .. controls (220.52,149.03) and (222.48,150.98) .. (222.48,153.39) .. controls (222.48,155.8) and (220.52,157.75) .. (218.1,157.75) .. controls (215.68,157.75) and (213.72,155.8) .. (213.72,153.39) -- cycle ;
		\draw    (41.32,213.65) -- (130.97,213.15) ;
		\draw    (59.85,111.34) -- (130.45,111.1) ;
		\draw    (130.45,111.1) -- (218.27,88.95) ;
		\draw    (130.97,213.15) -- (218.27,88.95) ;
		\draw  [line width=1.5]  (90.69,98.71) -- (94.49,100.83) -- (90.69,102.95) ;
		\draw  [line width=1.5]  (90.69,108.97) -- (94.49,111.09) -- (90.69,113.21) ;
		\draw  [line width=1.5]  (103.77,205.51) -- (99.97,203.39) -- (103.77,201.27) ;
		\draw  [line width=1.5]  (104.28,215.77) -- (100.49,213.65) -- (104.28,211.53) ;
		\draw  [line width=1.5]  (202.2,130.43) -- (202.33,134.75) -- (198.55,132.61) ;
		\draw  [line width=1.5]  (182.57,95.54) -- (186.72,96.87) -- (183.42,99.69) ;
		\draw  [line width=1.5]  (201.6,175.59) -- (197.32,176.4) -- (198.85,172.35) ;
		\draw  [line width=1.5]  (163.47,170.31) -- (159.51,172.09) -- (160.04,167.79) ;
		\draw   (172.47,202.62) .. controls (172,201.28) and (171.71,199.55) .. (171.71,197.39) .. controls (171.71,186.62) and (178.93,186.62) .. (178.93,192.26) .. controls (178.93,197.9) and (171.71,197.9) .. (171.71,187.13) .. controls (171.71,176.36) and (178.93,176.36) .. (178.93,182) .. controls (178.93,187.64) and (171.71,187.64) .. (171.71,176.87) .. controls (171.71,166.11) and (178.93,166.11) .. (178.93,171.75) .. controls (178.93,177.39) and (171.71,177.39) .. (171.71,166.62) .. controls (171.71,155.85) and (178.93,155.85) .. (178.93,161.49) .. controls (178.93,167.13) and (171.71,167.13) .. (171.71,156.36) .. controls (171.71,145.59) and (178.93,145.59) .. (178.93,151.23) .. controls (178.93,156.88) and (171.71,156.88) .. (171.71,146.11) .. controls (171.71,135.34) and (178.93,135.34) .. (178.93,140.98) .. controls (178.93,146.62) and (171.71,146.62) .. (171.71,135.85) .. controls (171.71,125.08) and (178.93,125.08) .. (178.93,130.72) .. controls (178.93,136.36) and (171.71,136.36) .. (171.71,125.59) .. controls (171.71,114.83) and (178.93,114.83) .. (178.93,120.47) .. controls (178.93,126.11) and (171.71,126.11) .. (171.71,115.34) .. controls (171.71,104.57) and (178.93,104.57) .. (178.93,110.21) .. controls (178.93,115.85) and (171.71,115.85) .. (171.71,105.08) .. controls (171.71,102.75) and (172.05,100.92) .. (172.58,99.54) ;
		\draw   (131.67,213.15) .. controls (131.24,211.81) and (130.97,210.08) .. (130.97,207.92) .. controls (130.97,197.15) and (137.67,197.15) .. (137.67,202.79) .. controls (137.67,208.43) and (130.97,208.43) .. (130.97,197.66) .. controls (130.97,186.89) and (137.67,186.89) .. (137.67,192.53) .. controls (137.67,198.17) and (130.97,198.17) .. (130.97,187.41) .. controls (130.97,176.64) and (137.67,176.64) .. (137.67,182.28) .. controls (137.67,187.92) and (130.97,187.92) .. (130.97,177.15) .. controls (130.97,166.38) and (137.67,166.38) .. (137.67,172.02) .. controls (137.67,177.66) and (130.97,177.66) .. (130.97,166.89) .. controls (130.97,156.12) and (137.67,156.12) .. (137.67,161.77) .. controls (137.67,167.41) and (130.97,167.41) .. (130.97,156.64) .. controls (130.97,145.87) and (137.67,145.87) .. (137.67,151.51) .. controls (137.67,157.15) and (130.97,157.15) .. (130.97,146.38) .. controls (130.97,135.61) and (137.67,135.61) .. (137.67,141.25) .. controls (137.67,146.89) and (130.97,146.89) .. (130.97,136.13) .. controls (130.97,125.36) and (137.67,125.36) .. (137.67,131) .. controls (137.67,136.64) and (130.97,136.64) .. (130.97,125.87) .. controls (130.97,115.1) and (137.67,115.1) .. (137.67,120.74) .. controls (137.67,126.38) and (130.97,126.38) .. (130.97,115.61) .. controls (130.97,113.68) and (131.18,112.09) .. (131.54,110.82) ;
		\draw  [fill={rgb, 255:red, 74; green, 144; blue, 226 }  ,fill opacity=1 ] (213.89,89.3) .. controls (213.89,86.9) and (215.85,84.95) .. (218.27,84.95) .. controls (220.69,84.95) and (222.65,86.9) .. (222.65,89.3) .. controls (222.65,91.71) and (220.69,93.66) .. (218.27,93.66) .. controls (215.85,93.66) and (213.89,91.71) .. (213.89,89.3) -- cycle ;
		\draw  [fill={rgb, 255:red, 0; green, 0; blue, 0 }  ,fill opacity=0.69 ] (395.49,88.21) .. controls (409.42,88.19) and (420.74,115.59) .. (420.78,149.4) .. controls (420.82,183.21) and (409.55,210.63) .. (395.62,210.64) .. controls (381.68,210.66) and (370.36,183.26) .. (370.32,149.45) .. controls (370.29,115.64) and (381.56,88.22) .. (395.49,88.21) -- cycle ;
		\draw    (395.49,88.21) -- (496.11,87.31) ;
		\draw    (496.11,87.31) -- (570.51,210.56) ;
		\draw    (414.67,190.39) -- (495.6,190.39) ;
		\draw    (495.6,190.39) -- (570.51,210.56) ;
		\draw   (496.36,190.39) .. controls (495.89,189.05) and (495.6,187.32) .. (495.6,185.16) .. controls (495.6,174.39) and (502.82,174.39) .. (502.82,180.03) .. controls (502.82,185.67) and (495.6,185.67) .. (495.6,174.9) .. controls (495.6,164.13) and (502.82,164.13) .. (502.82,169.77) .. controls (502.82,175.41) and (495.6,175.41) .. (495.6,164.64) .. controls (495.6,153.88) and (502.82,153.88) .. (502.82,159.52) .. controls (502.82,165.16) and (495.6,165.16) .. (495.6,154.39) .. controls (495.6,143.62) and (502.82,143.62) .. (502.82,149.26) .. controls (502.82,154.9) and (495.6,154.9) .. (495.6,144.13) .. controls (495.6,133.36) and (502.82,133.36) .. (502.82,139) .. controls (502.82,144.64) and (495.6,144.64) .. (495.6,133.88) .. controls (495.6,123.11) and (502.82,123.11) .. (502.82,128.75) .. controls (502.82,134.39) and (495.6,134.39) .. (495.6,123.62) .. controls (495.6,112.85) and (502.82,112.85) .. (502.82,118.49) .. controls (502.82,124.13) and (495.6,124.13) .. (495.6,113.36) .. controls (495.6,102.59) and (502.82,102.59) .. (502.82,108.24) .. controls (502.82,113.88) and (495.6,113.88) .. (495.6,103.11) .. controls (495.6,92.34) and (502.82,92.34) .. (502.82,97.98) .. controls (502.82,103.62) and (495.6,103.62) .. (495.6,92.85) .. controls (495.6,90.52) and (495.94,88.69) .. (496.47,87.31) ;
		\draw  [fill={rgb, 255:red, 139; green, 87; blue, 42 }  ,fill opacity=1 ] (566.13,210.56) .. controls (566.13,208.15) and (568.09,206.2) .. (570.51,206.2) .. controls (572.93,206.2) and (574.89,208.15) .. (574.89,210.56) .. controls (574.89,212.96) and (572.93,214.92) .. (570.51,214.92) .. controls (568.09,214.92) and (566.13,212.96) .. (566.13,210.56) -- cycle ;
		\draw  [line width=1.5]  (444.48,85.45) -- (448.27,87.57) -- (444.48,89.69) ;
		\draw  [line width=1.5]  (511.93,108.84) -- (512.07,113.16) -- (508.28,111.02) ;
		\draw  [line width=1.5]  (546.67,205.98) -- (543.5,203.01) -- (547.7,201.87) ;
		\draw  [line width=1.5]  (458.07,192.26) -- (454.27,190.14) -- (458.07,188.01) ;
		\draw    (409.52,98.08) -- (515.91,96.52) ;
		\draw    (515.91,96.52) -- (573.91,78.52) ;
		\draw    (410.03,200.13) -- (487.91,200.52) ;
		\draw    (487.91,200.52) -- (572.4,150.39) ;
		\draw  [fill={rgb, 255:red, 74; green, 144; blue, 226 }  ,fill opacity=1 ] (568.02,150.39) .. controls (568.02,147.98) and (569.98,146.03) .. (572.4,146.03) .. controls (574.82,146.03) and (576.79,147.98) .. (576.79,150.39) .. controls (576.79,152.8) and (574.82,154.75) .. (572.4,154.75) .. controls (569.98,154.75) and (568.02,152.8) .. (568.02,150.39) -- cycle ;
		\draw    (395.62,210.64) -- (514.33,211.33) ;
		\draw    (414.16,108.34) -- (484.75,108.1) ;
		\draw    (484.75,108.1) -- (520.33,126.33) -- (568.02,150.39) ;
		\draw    (514.33,211.33) -- (573.91,78.52) ;
		\draw  [line width=1.5]  (444.99,95.71) -- (448.79,97.83) -- (444.99,99.95) ;
		\draw  [line width=1.5]  (444.99,105.97) -- (448.79,108.09) -- (444.99,110.21) ;
		\draw  [line width=1.5]  (458.07,202.51) -- (454.27,200.39) -- (458.07,198.27) ;
		\draw  [line width=1.5]  (458.58,211.77) -- (454.79,209.65) -- (458.58,207.53) ;
		\draw  [line width=1.5]  (549.88,138.41) -- (552.36,141.95) -- (548.02,142.24) ;
		\draw  [line width=1.5]  (537.6,87.99) -- (541.92,88.53) -- (539.2,91.92) ;
		\draw  [line width=1.5]  (548.28,169.6) -- (547.82,173.93) -- (544.38,171.29) ;
		\draw  [line width=1.5]  (545.71,147.92) -- (542.03,150.23) -- (541.97,145.9) ;
		\draw   (515.28,212.33) .. controls (514.7,211) and (514.33,209.27) .. (514.33,207.1) .. controls (514.33,196.33) and (523.33,196.33) .. (523.33,201.97) .. controls (523.33,207.62) and (514.33,207.62) .. (514.33,196.85) .. controls (514.33,186.08) and (523.33,186.08) .. (523.33,191.72) .. controls (523.33,197.36) and (514.33,197.36) .. (514.33,186.59) .. controls (514.33,175.82) and (523.33,175.82) .. (523.33,181.46) .. controls (523.33,187.1) and (514.33,187.1) .. (514.33,176.33) .. controls (514.33,165.57) and (523.33,165.57) .. (523.33,171.21) .. controls (523.33,176.85) and (514.33,176.85) .. (514.33,166.08) .. controls (514.33,155.31) and (523.33,155.31) .. (523.33,160.95) .. controls (523.33,166.59) and (514.33,166.59) .. (514.33,155.82) .. controls (514.33,145.05) and (523.33,145.05) .. (523.33,150.69) .. controls (523.33,156.34) and (514.33,156.34) .. (514.33,145.57) .. controls (514.33,134.8) and (523.33,134.8) .. (523.33,140.44) .. controls (523.33,146.08) and (514.33,146.08) .. (514.33,135.31) .. controls (514.33,124.54) and (523.33,124.54) .. (523.33,130.18) .. controls (523.33,135.82) and (514.33,135.82) .. (514.33,125.05) .. controls (514.33,114.29) and (523.33,114.29) .. (523.33,119.93) .. controls (523.33,125.57) and (514.33,125.57) .. (514.33,114.8) .. controls (514.33,104.03) and (523.33,104.03) .. (523.33,109.67) .. controls (523.33,115.31) and (514.33,115.31) .. (514.33,104.54) .. controls (514.33,93.77) and (523.33,93.77) .. (523.33,99.41) .. controls (523.33,104.83) and (515.05,105.05) .. (514.38,95.54) ;
		\draw   (486.18,200.52) .. controls (485.62,199.18) and (485.27,197.45) .. (485.27,195.29) .. controls (485.27,184.52) and (493.91,184.52) .. (493.91,190.16) .. controls (493.91,195.8) and (485.27,195.8) .. (485.27,185.03) .. controls (485.27,174.26) and (493.91,174.26) .. (493.91,179.9) .. controls (493.91,185.54) and (485.27,185.54) .. (485.27,174.77) .. controls (485.27,164.01) and (493.91,164.01) .. (493.91,169.65) .. controls (493.91,175.29) and (485.27,175.29) .. (485.27,164.52) .. controls (485.27,153.75) and (493.91,153.75) .. (493.91,159.39) .. controls (493.91,165.03) and (485.27,165.03) .. (485.27,154.26) .. controls (485.27,143.49) and (493.91,143.49) .. (493.91,149.13) .. controls (493.91,154.77) and (485.27,154.77) .. (485.27,144.01) .. controls (485.27,133.24) and (493.91,133.24) .. (493.91,138.88) .. controls (493.91,144.52) and (485.27,144.52) .. (485.27,133.75) .. controls (485.27,122.98) and (493.91,122.98) .. (493.91,128.62) .. controls (493.91,134.26) and (485.27,134.26) .. (485.27,123.49) .. controls (485.27,112.73) and (493.91,112.73) .. (493.91,118.37) .. controls (493.91,124.01) and (485.27,124.01) .. (485.27,113.24) .. controls (485.27,110.97) and (485.65,109.18) .. (486.26,107.82) ;
		\draw  [fill={rgb, 255:red, 139; green, 87; blue, 42 }  ,fill opacity=1 ] (569.53,78.52) .. controls (569.53,76.11) and (571.49,74.16) .. (573.91,74.16) .. controls (576.33,74.16) and (578.3,76.11) .. (578.3,78.52) .. controls (578.3,80.92) and (576.33,82.88) .. (573.91,82.88) .. controls (571.49,82.88) and (569.53,80.92) .. (569.53,78.52) -- cycle ;
		
		\draw (3.26,84.11) node [anchor=north west][inner sep=0.75pt]    {$\Lambda _{Q}$};
		\draw (-1.3,202.71) node [anchor=north west][inner sep=0.75pt]    {$\bar{\Sigma }_{Q}$};
		\draw (224.17,205.79) node [anchor=north west][inner sep=0.75pt]    {$\pi $};
		\draw (225.53,140.01) node [anchor=north west][inner sep=0.75pt]    {$B\left( B^{*}\right) /\bar{D}\left( \bar{D}^{*}\right)$};
		\draw (224.34,72.47) node [anchor=north west][inner sep=0.75pt]    {$\bar{B}\left( \bar{B}^{*}\right) /D\left( D^{*}\right)$};
		\draw (357.56,81.11) node [anchor=north west][inner sep=0.75pt]    {$\Lambda _{Q}$};
		\draw (353,199.71) node [anchor=north west][inner sep=0.75pt]    {$\bar{\Sigma }_{Q}$};
		\draw (573.47,202.79) node [anchor=north west][inner sep=0.75pt]    {$\pi $};
		\draw (582.47,64.79) node [anchor=north west][inner sep=0.75pt]    {$\pi $};
		\draw (563.34,156.00) node [anchor=north west][inner sep=0.75pt]    {$\left( ^{1} S_{0} \ \text{or } ^{3} S_{1}\right)$};
		\draw (589,132.4) node [anchor=north west][inner sep=0.75pt]    {$Q\bar{Q}$};

	\end{tikzpicture}
	\caption{The mechanism of the three-body strong decay modes of $\Lambda_Q \bar{\Sigma}_Q$}
	\label{fig:strong}
\end{figure}
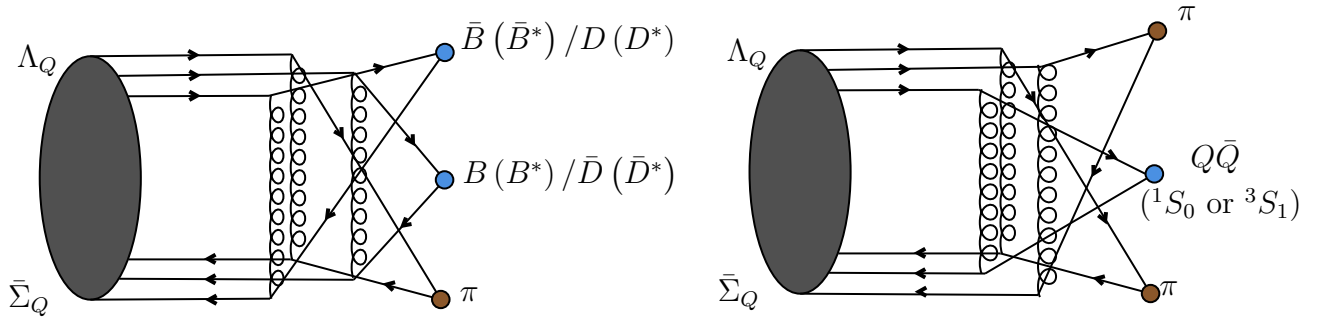

For the $\Lambda_Q \bar{\Sigma}_Q$ baryonium states, the dominant decay mode proceeds via strong interactions directly into the corresponding baryons, $\Lambda_Q$ and $\bar{\Sigma}_Q$.  Since the masses of the $\Lambda_Q \bar{\Sigma}_Q$ states all lie above their respective physical thresholds, the mode to the corresponding baryon–antibaryon pairs are allowed. Another dominant decay mode is the three-body strong decay, in which the final states consist of three mesons. The corresponding decay mechanism is illustrated in Fig.~\ref{fig:strong}. In addition, weak decays are also possible. However, they involve Cabibbo-suppressed processes and therefore do not constitute the dominant decay modes. The dominant strong decay channels of the $\Lambda_Q \bar{\Sigma}_Q$ states are summarized in Table.~\ref{Decay}.

\begin{table}[h]
		\linespread{1.3}\selectfont
		\centering
		\begin{tabular}{ccccc}
			\hline
			\hline
			$J^{P}$ & $0^{-}$ & $0^{+}$ & $1^{-}$ & $1^+$ \\
			\hline
			$\Lambda_c \bar{\Sigma}_c$
			& $\Lambda_c\; \bar{\Sigma}_c$, $\pi\pi \eta_c$ & $\Lambda_c\; \bar{\Sigma}_c$  & $\Lambda_c\; \bar{\Sigma}_c$, $\pi\pi J/\psi$ & $\Lambda_c\; \bar{\Sigma}_c$ \\
			& $\pi D \bar{D}$, $\pi D^* \bar{D}^*$ &   &$\pi D^* \bar{D}^*$, $\pi D \bar{D}^*$ &     \\
			\hline
			$\Lambda_b \bar{\Sigma}_b$
			& $\Lambda_b\; \bar{\Sigma}_b$, $\pi\pi \eta_b$ & $\Lambda_b\; \bar{\Sigma}_b$ & $\Lambda_b\; \bar{\Sigma}_b$, $\pi\pi \Upsilon$ & $\Lambda_b\; \bar{\Sigma}_b$ \\
			& $\pi B \bar{B}$, $\pi B^*\bar{B}^*$  &   &$\pi B^*\bar{B}^*$, $\pi B \bar{B}^*$ &  \\
			\hline\hline
		\end{tabular}
	\caption{Typical decay modes of the $\Lambda_Q \bar{\Sigma}_Q$ states. Besides, if the non-zero relative orbital angular momenta between the decay products are taken into account, the states with $J^P = 0^+, 1^+$ can also possess decay modes whose final states consist of three mesons.
	}
	\label{Decay}
\end{table}

\section{Discussion and Conclusions\label{dc}}
In summary, in this work we have employed the QCDSR method to calculate the masses and decay constants of the ground hidden-charm and hidden-bottom $\Lambda_Q \bar{\Sigma}_Q$ states. Two linearly independent hadronic interpolating currents are constructed for this purpose, and nonperturbative contribution up to dimension 12 is considered. Possible decay modes of these states have also been analyzed. The results indicate that the baryonium currents constructed from the Type-I baryonic currents yield reliable Borel windows for the $\Lambda_b \bar{\Sigma}_b$ states with $J^P = 0^-$ and $1^-$. In contrast, the currents constructed from the Type-II baryonic currents provide reliable Borel windows for all four quantum-number assignments $J^P = 0^-, 0^+, 1^-, 1^+$ of the $\Lambda_Q \bar{\Sigma}_Q$ ground states. All extracted masses lie above the corresponding baryon–antibaryon thresholds.

In our calculation, the central value of the $\Lambda_c \bar{\Sigma}_c$ state lies in the range $5.7$--$5.8$~GeV, more than $1$~GeV above its baryon–antibaryon threshold $E_{\mathrm{th},c}=4.71~\mathrm{GeV}$, indicating that the extracted structure is unlikely to correspond to a near-threshold bound state. The absence of any bound state signal near threshold in BESIII measurements \cite{BESIII:2025zgc} is therefore consistent with our theoretical interpretation. The fact that the present result does not contradict existing experimental observations supports the reliability of the QCDSR method at the qualitative level. In particular, although the typical uncertainty is of order $\Lambda_{\mathrm{QCD}}$, it remains sufficient to distinguish between near-threshold configurations and states located well above the corresponding hadronic threshold. At the same time, the detailed theoretical understanding of the applicability of the QCD sum rule approach to multiquark systems remains to be further explored. In particular, clarifying the interplay between the OPE and the complex spectral structure of such systems would be an interesting direction for future studies. 

Theoretical predictions \cite{Dong:2021juy} cited in experimental studies differ from our results, likely due to different binding mechanisms. In the Bethe-Salpeter approach, the baryon–antibaryon interaction arises from single-meson exchange at the hadronic level, yielding near-threshold masses that depend on the potential model. In contrast, QCD sum rules probe binding at the quark level via gluon and mixed condensates, which connect to hadronic observables through quark-hadron duality. Moreover, the molecular nature is reflected in the color structure of the interpolating currents ($[1]_c-[1]_c$), which dictates the configuration of condensate contributions. Consequently, QCDSR results depend sensitively on the choice of interpolating currents. A fully reliable determination of the mass of a baryon–antibaryon molecular or resonance-like state, as well as deeper insight into the nonperturbative hadronization process, requires further study.

It should also be noted that Ref.~\cite{Chen:2016ymy} systematically studied hidden-charm hexaquark states. In their results, only the $J^P = 0^+, 1^+$ hidden-charm baryonium states were found, with some of their masses falling in the range of $5.5$--$6.0$~GeV, which is consistent with our findings. The main differences between that work and the present study are twofold: First, the $\Sigma$ baryon interpolating currents used in the two works are different, leading to the construction of distinct baryonium currents that could couple to different states, which is why the $J^P = 0^-, 1^-$ hidden-charm baryonium states $\Lambda_c \bar{\Sigma}_c$ are found in this work and obtained different mass values. Second, this work takes into account a larger number of nonperturbative condensate contributions, such as the dimension-10 condensate $\langle q\bar{q} \rangle^2 \langle G^2 \rangle$ and the dimension-11 condensate $\langle q\bar{q} \rangle^2 \langle qG\bar{q} \rangle$, which were not considered in their work. From our calculations, it is evident that the contribution of $\langle q\bar{q} \rangle^2 \langle qG\bar{q} \rangle$ is greater than $10\%$ in many states, and can even approach $20\%$. This has a significant impact, directly affecting the existence of $J^P = 0^-, 1^-$ states.

As is discussed above, to search for the $\Lambda_c \bar{\Sigma}_c$ state, experiments could increase the center-of-mass energy and explore the region $5.5$--$6.0$~GeV. For the hidden-bottom baryonium states $\Lambda_b\bar{\Sigma}_b$, no possible signals have been observed experimentally so far. The states we have calculated would be detected in experiments such as STCF, LHCb, ATLAS, BelleII, and others. They can serve as candidates for hidden-bottom resonances in the $12$~GeV  region.

\vspace{0.5cm}
{\bf Acknowledgments}

 Xuan-Heng Zhang appreciates Bing-Dong Wan and Liang Tang for their highlight discussions. This work was supported in part by the National Key Research and Development Program of China under Contracts No.~2025YFA1613900, by the National Natural Science Foundation of China(NSFC) under the Grants 12475087 and 12235008. 

\bibliographystyle{JHEP}
\bibliography{references.bib}

\newpage

\begin{appendix}
	In the appendix, the analytical results for the spectral densities are presented, corresponding to 8 different configurations, i.e., four quantum numbers for each of the hexaquark states with two different baryonic currents. In calculating the spectral densities, the \texttt{FeynCalc} package \cite{Shtabovenko:2020gxv,Shtabovenko:2016sxi,Mertig:1990an} is utilized to trace out the $\gamma$-matrices. 
	
	We expand the spectral densities as Eq.~\eqref{OPE}, the zero-contribution term will not be displayed. The dynamical quantities are defined as \cite{Albuquerque:2012jbz, Wan:2019ake}
	\begin{equation}
		F_{\alpha\beta}=(\alpha+\beta)m_Q^2-\alpha\beta s;\quad H_{\alpha}=m_Q^2-\alpha(1-\alpha)s,
	\end{equation}
	where $\alpha,\beta$ are integration variables. Their integration limits are defined as
	\begin{equation}
		\alpha_{\min}=\frac{1}{2}\left(1-v\right),\quad \alpha_{\max}=\frac{1}{2}\left(1+v\right);
	\end{equation}
	\begin{equation}
		\beta_{\min}=\frac{\alpha m_Q^2}{s\alpha-m_Q^2},
	\end{equation}
	where $v=\sqrt{1-\dfrac{4m_Q^2}{s}}$. Therefore, the integrate measure is defined as
	\begin{equation}
		\int_{\alpha}=\int_{\alpha_{\min}}^{\alpha_{\max}}\dd\alpha.\quad \int_{\beta}=\int_{\beta_{\min}}^{1-\alpha}\dd\beta\ .
	\end{equation}
	
	\section{Type-I Currents}
	
	\subsection{$0^-$ $\Lambda_Q\bar{\Sigma}_Q$ States}
	\begin{eqnarray}
		\rho^{\text{pert}}&=& - \int_{\alpha}\int_{\beta} \frac{F_{\alpha\beta}^7 (\alpha +\beta -1)^4}{55050240 \pi ^{10} \alpha ^6 \beta ^6},\\	
		\rho^{\expval{q\bar{q}}}&=&-\int_{\alpha}\int_{\beta}\frac{F_{\alpha\beta}^5 m_Q \expval{q\bar{q}} (\alpha -\beta ) (\alpha +\beta -1)^3}{245760 \pi ^8 \alpha ^5 \beta ^5},\\
		\rho^{\expval{G^2}}&=& -\expval{g_s^2G^2}\int_{\alpha}\int_{\beta}\left[\frac{F_{\alpha\beta}^5 (2 \alpha -\beta +1) (\alpha +\beta -1)^2}{15728640 \pi ^{10} \alpha ^5 \beta ^4}\right.\\
		& &\left.+\frac{F_{\alpha\beta}^4 m_Q^2  (\alpha +\beta -1)^4 \left(\alpha ^3+\beta ^3\right)}{18874368 \pi ^{10} \alpha ^6 \beta ^6}\right],\\
		\rho^{\expval{qG\bar{q}}}&=&m_Q\expval{qG\bar{q}}\int_{\alpha}\int_{\beta}\left[\frac{F_{\alpha\beta}^4  (\alpha -\beta ) (\alpha +\beta -1)^2}{65536 \pi ^8 \alpha ^4 \beta ^4}-\frac{F_{\alpha\beta}^4  (\alpha +\beta -1)^3}{196608 \pi ^8 \alpha ^5 \beta ^3}\right],\\
		\rho^{\expval{q\bar{q}}^2}&=&\expval{q\bar{q}}^2\int_{\alpha}\int_{\beta}\frac{F_{\alpha\beta}^3 (\alpha +\beta -1) \left(F_{\alpha\beta}+2 m_Q^2 (\alpha +\beta -1)\right)}{6144 \pi ^6 \alpha ^3 \beta ^3},\\
		\rho^{\expval{G^3}}&=&-\expval{g_s^3G^3}\int_{\alpha}\int_{\beta}\frac{F_{\alpha\beta}^3 (\alpha +\beta -1)^4 \left(F_{\alpha\beta} \left(\alpha ^3+\beta ^3\right)+8 m_Q^2 \left(\alpha ^4+\beta ^4\right)\right)}{75497472 \pi ^{10} \alpha ^6 \beta ^6},\\
		\rho^{\expval{q\bar{q}}\expval{G^2}}&=&m_Q\expval{q\bar{q}}\expval{g_s^2G^2}\int_{\alpha}\int_{\beta}\left[-\frac{F_{\alpha\beta}^3  (\alpha -3 \beta +1) (\alpha +\beta -1)}{196608 \pi ^8 \alpha ^3 \beta ^3}\right.\\
		& &\left.+\frac{F_{\alpha\beta}^2 (\beta-\alpha) (\alpha +\beta -1)^3 \left(F_{\alpha\beta} \left(\alpha ^2+\alpha  \beta +\beta ^2\right)+m_Q^2 \left(\alpha ^3+\beta ^3\right)\right)}{294912 \pi ^8 \alpha ^5 \beta ^5}\right],\\
		\rho^{\expval{q\bar{q}}\expval{qG\bar{q}}}&=&\expval{q\bar{q}}\expval{qG\bar{q}}\int_{\alpha}\int_{\beta}\left[\frac{F_{\alpha\beta}^2 m_Q^2  (-\alpha -\beta +1)}{1024 \pi ^6 \alpha ^2 \beta ^2}-\frac{F_{\alpha\beta}^3}{3072 \pi ^6 \alpha ^2 \beta ^2}\right],\\
		\rho^{\expval{q\bar{q}}^3}&=&\int_{\alpha}\int_{\beta}\frac{F_{\alpha\beta}^2 m_Q \expval{q\bar{q}}^3 (\alpha -\beta )}{384 \pi ^4 \alpha ^2 \beta ^2},\\
		\rho^{\expval{qG\bar{q}}\expval{G^2}}&=& m_Q\expval{qG\bar{q}}\expval{G^2}\int_{\alpha}\int_{\beta}\left[\frac{3F^2_{\alpha\beta} (\alpha -\beta ) (\alpha +\beta -1)^2 \left(\alpha ^2+\alpha  \beta +\beta ^2\right)}{393216 \pi ^8 \alpha ^4 \beta ^4}\right.\\
		& &\left.+\frac{F_{\alpha\beta}(\alpha -\beta ) (\alpha +\beta -1)^2 m_Q^2 \left(\alpha ^3+\beta ^3\right)}{196608 \pi ^8 \alpha ^4 \beta ^4}-\frac{(2 \beta -1) F_{\alpha\beta}^2 }{131072 \pi ^8 \alpha ^2 \beta ^2}\right],\\
		\rho^{\expval{qG\bar{q}}^2}&=&\expval{qG\bar{q}}^2\int_{\alpha}\left[ -\frac{H_{\alpha}^2 }{8192 \pi ^6 (1-\alpha ) \alpha }+\int_{\beta}\frac{F_{\alpha\beta}m_Q^2 }{4096 \pi ^6 \alpha  \beta} \right],\\
		\rho^{\expval{q\bar{q}}^2\expval{G^2}}&=&-\expval{q\bar{q}}^2\expval{G^2}\int_{\alpha}\int_{\beta}\left[\frac{(\alpha +\beta -1)m_Q^4 \left(\alpha ^4+\alpha ^3 (\beta -1)+\alpha  \beta ^3+(\beta -1) \beta ^3\right)}{36864 \pi ^6 \alpha ^3 \beta ^3}\right.\\
		& &\left.+\frac{ (\alpha +\beta -1) F_{\alpha\beta} m_Q^2 \left(5 \alpha ^3+3 \alpha ^2 (\beta -1)+3 \alpha  \beta ^2+(5 \beta -3) \beta ^2\right)}{36864 \pi ^6 \alpha ^3 \beta ^3}    \right.,\\
		& & \left.+\frac{F_{\alpha\beta}  \left(F_{\alpha\beta}-2 \alpha  m_Q^2\right)}{24576 \pi ^6 \alpha ^2 \beta }\right],\\
		\rho^{\expval{q\bar{q}}^4}&=&\int_{\alpha}\frac{m_Q^2 \expval{q\bar{q}}^4}{144 \pi ^2}.
	\end{eqnarray}
	
	\subsection{$0^+$ $\Lambda_Q\bar{\Sigma}_Q$ States}
	\begin{eqnarray}
		\rho^{\text{pert}}&=& - \int_{\alpha}\int_{\beta} \frac{F_{\alpha\beta}^7 (\alpha +\beta -1)^4}{55050240 \pi ^{10} \alpha ^6 \beta ^6},\\	
		\rho^{\expval{q\bar{q}}}&=&\int_{\alpha}\int_{\beta}\frac{F_{\alpha\beta}^5 m_Q \expval{q\bar{q}} (\alpha -\beta ) (\alpha +\beta -1)^3}{245760 \pi ^8 \alpha ^5 \beta ^5},\\
		\rho^{\expval{G^2}}&=& -\expval{g_s^2G^2}\int_{\alpha}\int_{\beta}\left[\frac{F_{\alpha\beta}^5 (2 \alpha -\beta +1) (\alpha +\beta -1)^2}{15728640 \pi ^{10} \alpha ^5 \beta ^4}\right.\\
		& &\left.+\frac{F_{\alpha\beta}^4 m_Q^2  (\alpha +\beta -1)^4 \left(\alpha ^3+\beta ^3\right)}{18874368 \pi ^{10} \alpha ^6 \beta ^6}\right],\\
		\rho^{\expval{qG\bar{q}}}&=&-m_Q\expval{qG\bar{q}}\int_{\alpha}\int_{\beta}\left[\frac{F_{\alpha\beta}^4  (\alpha +\beta ) (\alpha +\beta -1)^2}{65536 \pi ^8 \alpha ^4 \beta ^4}+\frac{F_{\alpha\beta}^4  (\alpha +\beta -1)^3}{196608 \pi ^8 \alpha ^5 \beta ^3}\right],\\
		\rho^{\expval{q\bar{q}}^2}&=&\expval{q\bar{q}}^2\int_{\alpha}\int_{\beta}\frac{F_{\alpha\beta}^3 (\alpha +\beta -1) \left(F_{\alpha\beta}-2 m_Q^2 (\alpha +\beta -1)\right)}{6144 \pi ^6 \alpha ^3 \beta ^3},\\
		\rho^{\expval{G^3}}&=&-\expval{g_s^3G^3}\int_{\alpha}\int_{\beta}\frac{F_{\alpha\beta}^3 (\alpha +\beta -1)^4 \left(F_{\alpha\beta} \left(\alpha ^3+\beta ^3\right)+8 m_Q^2 \left(\alpha ^4+\beta ^4\right)\right)}{75497472 \pi ^{10} \alpha ^6 \beta ^6},\\
		\rho^{\expval{q\bar{q}}\expval{G^2}}&=&m_Q\expval{q\bar{q}}\expval{g_s^2G^2}\int_{\alpha}\int_{\beta}\left[-\frac{F_{\alpha\beta}^3  (\alpha + \beta -1) (\alpha +\beta +1)}{196608 \pi ^8 \alpha ^3 \beta ^3}\right.\\
		& & \left.+\frac{F_{\alpha\beta}^2 (\beta^3+\alpha^3) (\alpha +\beta -1)^3 \left(F_{\alpha\beta}+m_Q^2 \left(\alpha +\beta \right)\right)}{294912 \pi ^8 \alpha ^5 \beta ^5}\right],\\
		\rho^{\expval{q\bar{q}}\expval{qG\bar{q}}}&=&-\expval{q\bar{q}}\expval{qG\bar{q}}\int_{\alpha}\int_{\beta}\left[\frac{F_{\alpha\beta}^2 m_Q^2  (-\alpha -\beta +1)}{1024 \pi ^6 \alpha ^2 \beta ^2}+\frac{F_{\alpha\beta}^3}{3072 \pi ^6 \alpha ^2 \beta ^2}\right],\\
		\rho^{\expval{q\bar{q}}^3}&=&-\int_{\alpha}\int_{\beta}\frac{F_{\alpha\beta}^2 m_Q \expval{q\bar{q}}^3 (\alpha +\beta )}{384 \pi ^4 \alpha ^2 \beta ^2},\\
		\rho^{\expval{qG\bar{q}}\expval{G^2}}&=& m_Q\expval{qG\bar{q}}\expval{G^2}\int_{\alpha}\int_{\beta}\left[\frac{3F^2_{\alpha\beta}  (\alpha +\beta -1)^2 \left(\alpha ^3+\beta ^3\right)}{393216 \pi ^8 \alpha ^4 \beta ^4}-\frac{ F_{\alpha\beta}^2 }{131072 \pi ^8 \alpha ^2 \beta ^2}\right.\\
		& &\left.+\frac{F_{\alpha\beta}(\alpha +\beta ) (\alpha +\beta -1)^2 m_Q^2 \left(\alpha ^3+\beta ^3\right)}{196608 \pi ^8 \alpha ^4 \beta ^4}\right],\\
		\rho^{\expval{qG\bar{q}}^2}&=&\expval{qG\bar{q}}^2\int_{\alpha}\left[ \frac{H_{\alpha}^2 }{8192 \pi ^6 (1-\alpha ) \alpha }+\int_{\beta}\frac{F_{\alpha\beta}m_Q^2 }{4096 \pi ^6 \alpha  \beta} \right],\\
		\rho^{\expval{q\bar{q}}^2\expval{G^2}}&=&-\expval{q\bar{q}}^2\expval{G^2}\int_{\alpha}\int_{\beta}\left[\frac{(\alpha +\beta -1)m_Q^4 \left(\alpha ^4+\alpha ^3 (\beta -1)+\alpha  \beta ^3+(\beta -1) \beta ^3\right)}{36864 \pi ^6 \alpha ^3 \beta ^3} \right.\\
		& &\left. +\frac{ (\alpha +\beta -1) F_{\alpha\beta} m_Q^2 \left( \alpha ^3+3 \alpha ^2 (\beta -1)+3 \alpha  \beta ^2+(\beta -3) \beta ^2\right)}{36864 \pi ^6 \alpha ^3 \beta ^3}\right.\\
		& &\left.-\frac{F_{\alpha\beta}  \left(F_{\alpha\beta}+2 \alpha  m_Q^2\right)}{24576 \pi ^6 \alpha ^2 \beta }   \right],\\
		\rho^{\expval{q\bar{q}}^2\expval{qG\bar{q}}}&=&-\expval{q\bar{q}}^2\expval{qG\bar{q}}\int_{\alpha}\left[\frac{H_{\alpha} m_Q}{256 \pi ^4 (1-\alpha )}+\frac{H_{\alpha} m_Q }{256 \pi ^4 \alpha }\right],\\
		\rho^{\expval{q\bar{q}}^4}&=&-\int_{\alpha}\frac{m_Q^2 \expval{q\bar{q}}^4}{144 \pi ^2}.
	\end{eqnarray}
	
	\subsection{$1^-$ $\Lambda_Q\bar{\Sigma}_Q$ States}
	\begin{eqnarray}
		\rho^{\text{pert}}&=& - \int_{\alpha}\int_{\beta} \frac{F_{\alpha\beta}^7 (\alpha +\beta -1)^4(\alpha +\beta +4)}{275251200  \pi ^{10} \alpha ^6 \beta ^6},\\	
		\rho^{\expval{q\bar{q}}}&=&-\int_{\alpha}\int_{\beta}\frac{F_{\alpha\beta}^5 m_Q \expval{q\bar{q}} (\alpha  (\beta -4)+(\beta +3) \beta) (\alpha +\beta -1)^3}{983040 \pi ^8 \alpha ^5 \beta ^5},\\
		\rho^{\expval{G^2}}&=& -\expval{g_s^2G^2}\int_{\alpha}\int_{\beta}\left[\frac{F_{\alpha\beta}^5 (3 \alpha ^2+2 \alpha  (\beta +3)-\beta ^2-2 \beta +3) (\alpha +\beta -1)^2}{62914560 \pi ^{10} \alpha ^5 \beta ^4}\right.\\
		& &\left.+\frac{F_{\alpha\beta}^4 m_Q^2  (\alpha +\beta -1)^4 (\alpha ^4+\alpha ^3 (\beta +4)+\alpha  \beta ^3+(\beta +4) \beta ^3)}{94371840 \pi ^{10} \alpha ^6 \beta ^6}\right],\\
		\rho^{\expval{qG\bar{q}}}&=&-m_Q\expval{qG\bar{q}}\int_{\alpha}\int_{\beta}\left[\frac{F_{\alpha\beta}^4  (\alpha +\beta -1)^2(\alpha  (\beta -3)+(\beta +2) \beta)}{196608 \pi ^8 \alpha ^4 \beta ^4}\right.\\
		& &\left.+\frac{F_{\alpha\beta}^4 ( \alpha +\beta +3)(\alpha +\beta -1)^3}{786432 \pi ^8 \alpha ^5 \beta ^3}\right],\\
		\rho^{\expval{q\bar{q}}^2}&=&\expval{q\bar{q}}^2\int_{\alpha}\int_{\beta}\frac{F_{\alpha\beta}^3 (\alpha +\beta -1) \left(F_{\alpha\beta}(\alpha +\beta +1)+4 m_Q^2 (\alpha +\beta -1)\right)}{12288 \pi ^6 \alpha ^3 \beta ^3},\\
		\rho^{\expval{G^3}}&=&-\expval{g_s^3G^3}\int_{\alpha}\int_{\beta}\frac{F_{\alpha\beta}^3 (\alpha +\beta -1)^4(\alpha +\beta +4) \left(F_{\alpha\beta} \left(\alpha ^3+\beta ^3\right)+8 m_Q^2 \left(\alpha ^4+\beta ^4\right)\right)}{75497472 \pi ^{10} \alpha ^6 \beta ^6},\\
		\rho^{\expval{q\bar{q}}\expval{G^2}}&=&m_Q\expval{q\bar{q}}\expval{g_s^2G^2}\int_{\alpha}\int_{\beta}\left[\frac{F_{\alpha\beta}^3  (\alpha + \beta -1) (\alpha  (\beta -1)+\beta ^2+2 \beta -1)}{196608 \pi ^8 \alpha ^3 \beta ^3}\right.\\
		& & \left.+\frac{F_{\alpha\beta}^3 (\alpha +\beta -1)^3(-4 \alpha ^3+\alpha  \beta ^3+(\beta +3) \beta ^3) }{1179648 \pi ^8 \alpha ^5 \beta ^5}\right.\\
		& &\left.+\frac{F_{\alpha\beta}^2 m_Q^2 (\alpha ^4 (\beta -4)+\alpha ^3 (\beta +3) \beta +\alpha  (\beta -4) \beta ^3+(\beta +3) \beta ^4)}{1179648 \pi ^8 \alpha ^5 \beta ^5}\right],\\
		\rho^{\expval{q\bar{q}}\expval{qG\bar{q}}}&=&\expval{q\bar{q}}\expval{qG\bar{q}}\int_{\alpha}\int_{\beta}\left[\frac{F_{\alpha\beta}^2 m_Q^2  (-\alpha -\beta +1)}{1024 \pi ^6 \alpha ^2 \beta ^2}+\frac{F_{\alpha\beta}^3(-\alpha -\beta +1)}{3072 \pi ^6 \alpha ^2 \beta ^2}-\frac{F_{\alpha\beta}^3}{3072 \pi ^6 \alpha ^2 \beta ^2}\right],\\
		\rho^{\expval{q\bar{q}}^3}&=&-\int_{\alpha}\int_{\beta}\frac{F_{\alpha\beta}^2 m_Q \expval{q\bar{q}}^3 (\alpha  (\beta -1)+\beta ^2)}{384 \pi ^4 \alpha ^2 \beta ^2},\\
		\rho^{\expval{qG\bar{q}}\expval{G^2}}&=& m_Q\expval{qG\bar{q}}\expval{G^2}\int_{\alpha}\int_{\beta}\left[\frac{3F^2_{\alpha\beta} (-3 \alpha ^3+\alpha  \beta ^3+(\beta +2) \beta ^3 ) (\alpha +\beta -1)^2 }{1179648  \pi ^8 \alpha ^4 \beta ^4}\right.\\
		& &\left.+\frac{F_{\alpha\beta} (\alpha +\beta -1)^2 m_Q^2 \left(\alpha ^4 (\beta -3)+\alpha ^3 (\beta +2) \beta +\alpha  (\beta -3) \beta ^3+(\beta +2) \beta ^4\right)}{589824 \pi ^8 \alpha ^4 \beta ^4}\right.\\
		& &\left.-\frac{(\alpha  \beta +\beta ^2+\beta -1) F_{\alpha\beta}^2 }{131072 \pi ^8 \alpha ^2 \beta ^2}\right],\\
		\rho^{\expval{qG\bar{q}}^2}&=&\expval{qG\bar{q}}^2\int_{\alpha}\left\{-\frac{H_{\alpha}^2 }{8192 \pi ^6 (1-\alpha ) \alpha }+\int_{\beta}\left[\frac{F_{\alpha\beta}m_Q^2 }{4096 \pi ^6 \alpha  \beta} +\frac{F_{\alpha\beta}^2}{8192\pi ^6 \alpha  \beta}\right]\right\},\\
		\rho^{\expval{q\bar{q}}^2\expval{G^2}}&=&-\expval{q\bar{q}}^2\expval{G^2}\int_{\alpha}\int_{\beta}\left[\frac{(\alpha +\beta -1)m_Q^4 \left(\alpha ^4+\alpha ^3 (\beta -1)+\alpha  \beta ^3+(\beta -1) \beta ^3\right)}{36864 \pi ^6 \alpha ^3 \beta ^3}  \right.\\
		& +& \left.\frac{ (\alpha +\beta -1) F_{\alpha\beta} m_Q^2 \left(\alpha ^4+\alpha ^3 (\beta +4)+3 \alpha ^2 (\beta -1)+\alpha  (\beta +3) \beta ^2+\left(\beta ^2+4 \beta -3\right) \beta ^2\right)}{36864 \pi ^6 \alpha ^3 \beta ^3}\right.\\
		& &+\left.\frac{F_{\alpha\beta}  \left(F_{\alpha\beta}(\alpha+\beta)-2 \alpha  m_Q^2\right)}{24576 \pi ^6 \alpha ^2 \beta }\right],\\
		\rho^{\expval{q\bar{q}}^2\expval{qG\bar{q}}}&=&\expval{q\bar{q}}^2\expval{qG\bar{q}}\int_{\alpha}\int_{\beta}\frac{F_{\alpha\beta} m_Q}{256 \pi ^4 \alpha },\\
		\rho^{\expval{q\bar{q}}^4}&=&\int_{\alpha}\frac{m_Q^2 \expval{q\bar{q}}^4}{144 \pi ^2}.
	\end{eqnarray}
	
	\subsection{$1^+$ $\Lambda_Q\bar{\Sigma}_Q$ States}
	\begin{eqnarray}
		\rho^{\text{pert}}&=& - \int_{\alpha}\int_{\beta} \frac{F_{\alpha\beta}^7 (\alpha +\beta -1)^4(\alpha +\beta +4)}{275251200  \pi ^{10} \alpha ^6 \beta ^6},\\	
		\rho^{\expval{q\bar{q}}}&=&\int_{\alpha}\int_{\beta}\frac{F_{\alpha\beta}^5 m_Q \expval{q\bar{q}} (\alpha  (\beta +4)+(\beta +3) \beta) (\alpha +\beta -1)^3}{983040 \pi ^8 \alpha ^5 \beta ^5},\\
		\rho^{\expval{G^2}}&=& -\expval{g_s^2G^2}\int_{\alpha}\int_{\beta}\left[\frac{F_{\alpha\beta}^5 (3 \alpha ^2+2 \alpha  (\beta +3)-\beta ^2-2 \beta +3) (\alpha +\beta -1)^2}{62914560 \pi ^{10} \alpha ^5 \beta ^4}\right.\\
		& &\left.+\frac{F_{\alpha\beta}^4 m_Q^2  (\alpha +\beta -1)^4 (\alpha ^4+\alpha ^3 (\beta +4)+\alpha  \beta ^3+(\beta +4) \beta ^3)}{94371840 \pi ^{10} \alpha ^6 \beta ^6}\right],\\
		\rho^{\expval{qG\bar{q}}}&=&-m_Q\expval{qG\bar{q}}\int_{\alpha}\int_{\beta}\left[\frac{F_{\alpha\beta}^4  (\alpha +\beta -1)^2(\alpha  (\beta +3)+(\beta +2) \beta)}{196608 \pi ^8 \alpha ^4 \beta ^4}\right.\\
		& &\left.+\frac{F_{\alpha\beta}^4 ( \alpha +\beta +3)(\alpha +\beta -1)^3}{786432 \pi ^8 \alpha ^5 \beta ^3}\right],\\
		\rho^{\expval{q\bar{q}}^2}&=&\expval{q\bar{q}}^2\int_{\alpha}\int_{\beta}\frac{F_{\alpha\beta}^3 (\alpha +\beta -1) \left(F_{\alpha\beta}(\alpha +\beta +1)-4 m_Q^2 (\alpha +\beta -1)\right)}{12288 \pi ^6 \alpha ^3 \beta ^3},\\
		\rho^{\expval{G^3}}&=&-\expval{g_s^3G^3}\int_{\alpha}\int_{\beta}\frac{F_{\alpha\beta}^3 (\alpha +\beta -1)^4(\alpha+\beta+4) \left(F_{\alpha\beta} \left(\alpha ^3+\beta ^3\right)+8 m_Q^2 \left(\alpha ^4+\beta ^4\right)\right)}{377487360 \pi ^{10} \alpha ^6 \beta ^6},\\
		\rho^{\expval{q\bar{q}}\expval{G^2}}&=&m_Q\expval{q\bar{q}}\expval{g_s^2G^2}\int_{\alpha}\int_{\beta}\left[\frac{F_{\alpha\beta}^3  (\alpha + \beta -1) (\alpha  \beta +\alpha +\beta ^2+1)}{196608 \pi ^8 \alpha ^3 \beta ^3}\right.\\
		& & \left.+\frac{F_{\alpha\beta}^3 (\alpha +\beta -1)^3(4 \alpha ^3+\alpha  \beta ^3+(\beta +3) \beta ^3) }{1179648 \pi ^8 \alpha ^5 \beta ^5}\right.\\
		& &\left.+\frac{F_{\alpha\beta}^2 m_Q^2 (\alpha ^4 (\beta +4)+\alpha ^3 (\beta +3) \beta +\alpha  (\beta +4) \beta ^3+(\beta +3) \beta ^4)}{1179648 \pi ^8 \alpha ^5 \beta ^5}\right],\\
		\rho^{\expval{q\bar{q}}\expval{qG\bar{q}}}&=&\expval{q\bar{q}}\expval{qG\bar{q}}\int_{\alpha}\int_{\beta}\left[-\frac{F_{\alpha\beta}^2 m_Q^2  (-\alpha -\beta +1)}{1024 \pi ^6 \alpha ^2 \beta ^2}+\frac{F_{\alpha\beta}^3(-\alpha -\beta +1)}{3072 \pi ^6 \alpha ^2 \beta ^2}-\frac{F_{\alpha\beta}^3}{3072 \pi ^6 \alpha ^2 \beta ^2}\right],\\
		\rho^{\expval{q\bar{q}}^3}&=&-\int_{\alpha}\int_{\beta}\frac{F_{\alpha\beta}^2 m_Q \expval{q\bar{q}}^3 (\alpha  (\beta +1)+\beta ^2)}{384 \pi ^4 \alpha ^2 \beta ^2},\\
		\rho^{\expval{qG\bar{q}}\expval{G^2}}&=&- m_Q\expval{qG\bar{q}}\expval{G^2}\int_{\alpha}\int_{\beta}\left[\frac{3F^2_{\alpha\beta} (3 \alpha ^3+\alpha  \beta ^3+(\beta +2) \beta ^3 ) (\alpha +\beta -1)^2 }{1179648  \pi ^8 \alpha ^4 \beta ^4}\right.\\
		& &\left.+\frac{F_{\alpha\beta} (\alpha +\beta -1)^2 m_Q^2 \left(\alpha ^4 (\beta +3)+\alpha ^3 (\beta +2) \beta +\alpha  (\beta +3) \beta ^3+(\beta +2) \beta ^4\right)}{589824 \pi ^8 \alpha ^4 \beta ^4}\right.\\
		& &\left.+\frac{(\alpha  \beta +\beta ^2-\beta +1) F_{\alpha\beta}^2 }{131072 \pi ^8 \alpha ^2 \beta ^2}\right],\\
		\rho^{\expval{qG\bar{q}}^2}&=&\expval{qG\bar{q}}^2\int_{\alpha}\left\{-\frac{H_{\alpha}^2 }{8192 \pi ^6 (1-\alpha ) \alpha }+\int_{\beta}\left[\frac{F_{\alpha\beta}^2}{8192\pi ^6 \alpha  \beta}-\frac{F_{\alpha\beta}m_Q^2 }{4096 \pi ^6 \alpha  \beta} \right]\right\},\\
		\rho^{\expval{q\bar{q}}^2\expval{G^2}}&=&-\expval{q\bar{q}}^2\expval{G^2}\int_{\alpha}\int_{\beta}\left[\frac{(\alpha +\beta -1)m_Q^4 \left(\alpha ^4+\alpha ^3 (\beta -1)+\alpha  \beta ^3+(\beta -1) \beta ^3\right)}{36864 \pi ^6 \alpha ^3 \beta ^3}  \right.\\
		& +& \left.\frac{ (\alpha +\beta -1) F_{\alpha\beta} m_Q^2 \left(\alpha ^4+\alpha ^3 (\beta -2)-3 \alpha ^2 (\beta -1)+\alpha  (\beta -3) \beta ^2+\left(\beta ^2-2 \beta +3\right) \beta ^2\right)}{36864 \pi ^6 \alpha ^3 \beta ^3}\right.\\
		& &+\left.\frac{F_{\alpha\beta}  \left(F_{\alpha\beta}(\alpha+\beta)+2 \alpha  m_Q^2\right)}{24576 \pi ^6 \alpha ^2 \beta }\right],\\
		\rho^{\expval{q\bar{q}}^2\expval{qG\bar{q}}}&=&\expval{q\bar{q}}^2\expval{qG\bar{q}}\int_{\alpha}\left[-\frac{H_{\alpha} m_Q }{256 \pi ^4 (1-\alpha )}-\frac{H_{\alpha} m_Q }{256 \pi ^4 \alpha }+\int_{\beta}\frac{F_{\alpha\beta} m_Q}{256 \pi ^4 \alpha }\right],\\
		\rho^{\expval{q\bar{q}}^4}&=&-\int_{\alpha}\frac{m_Q^2 \expval{q\bar{q}}^4}{144 \pi ^2}.
	\end{eqnarray}

	\section{Type-II Currents}
	
	\subsection{$0^-$ $\Lambda_Q\bar{\Sigma}_Q$ States}
	\begin{eqnarray}
		\rho^{\text{pert}}&=& - \int_{\alpha}\int_{\beta} \frac{F_{\alpha\beta}^7 (\alpha +\beta -1)^4}{55050240 \pi ^{10} \alpha ^6 \beta ^6},\\	
		\rho^{\expval{q\bar{q}}}&=&-\int_{\alpha}\int_{\beta}\frac{F_{\alpha\beta}^5 m_Q \expval{q\bar{q}} (\alpha +\beta ) (\alpha +\beta -1)^3}{245760 \pi ^8 \alpha ^5 \beta ^5},\\
		\rho^{\expval{G^2}}&=& -\expval{g_s^2G^2}\int_{\alpha}\int_{\beta}\left[\frac{F_{\alpha\beta}^5 (2 \alpha -\beta +1) (\alpha +\beta -1)^2}{15728640 \pi ^{10} \alpha ^5 \beta ^4}\right.\\
		& &\left.+\frac{F_{\alpha\beta}^4 m_Q^2  (\alpha +\beta -1)^4 \left(\alpha ^3+\beta ^3\right)}{18874368 \pi ^{10} \alpha ^6 \beta ^6}\right],\\
		\rho^{\expval{qG\bar{q}}}&=&m_Q\expval{qG\bar{q}}\int_{\alpha}\int_{\beta}\left[\frac{F_{\alpha\beta}^4  (\alpha +\beta ) (\alpha +\beta -1)^2}{65536 \pi ^8 \alpha ^4 \beta ^4}+\frac{F_{\alpha\beta}^4  (\alpha +\beta -1)^3}{196608 \pi ^8 \alpha ^5 \beta ^3}\right],\\
		\rho^{\expval{q\bar{q}}^2}&=&-\expval{q\bar{q}}^2\int_{\alpha}\int_{\beta}\frac{F_{\alpha\beta}^3 (\alpha +\beta -1) \left(F_{\alpha\beta}+2 m_Q^2 (\alpha +\beta -1)\right)}{6144 \pi ^6 \alpha ^3 \beta ^3},\\
		\rho^{\expval{G^3}}&=&-\expval{g_s^3G^3}\int_{\alpha}\int_{\beta}\frac{F_{\alpha\beta}^3 (\alpha +\beta -1)^4 \left(F_{\alpha\beta} \left(\alpha ^3+\beta ^3\right)+8 m_Q^2 \left(\alpha ^4+\beta ^4\right)\right)}{75497472 \pi ^{10} \alpha ^6 \beta ^6},\\
		\rho^{\expval{q\bar{q}}\expval{G^2}}&=&-m_Q\expval{q\bar{q}}\expval{g_s^2G^2}\int_{\alpha}\int_{\beta}\left[\frac{F_{\alpha\beta}^3  (\alpha + \beta -1) (\alpha +\beta +1)}{196608 \pi ^8 \alpha ^3 \beta ^3}\right.\\
		& & \left.+\frac{F_{\alpha\beta}^2 (\beta^3+\alpha^3) (\alpha +\beta -1)^3 \left(F_{\alpha\beta}+m_Q^2 \left(\alpha +\beta \right)\right)}{294912 \pi ^8 \alpha ^5 \beta ^5}\right],\\
		\rho^{\expval{q\bar{q}}\expval{qG\bar{q}}}&=&-\expval{q\bar{q}}\expval{qG\bar{q}}\int_{\alpha}\int_{\beta}\left[\frac{F_{\alpha\beta}^2 m_Q^2  (-\alpha -\beta +1)}{1024 \pi ^6 \alpha ^2 \beta ^2}-\frac{F_{\alpha\beta}^3}{3072 \pi ^6 \alpha ^2 \beta ^2}\right],\\
		\rho^{\expval{q\bar{q}}^3}&=&-\int_{\alpha}\int_{\beta}\frac{F_{\alpha\beta}^2 m_Q \expval{q\bar{q}}^3 (\alpha +\beta )}{384 \pi ^4 \alpha ^2 \beta ^2},\\
		\rho^{\expval{qG\bar{q}}\expval{G^2}}&=& m_Q\expval{qG\bar{q}}\expval{G^2}\int_{\alpha}\int_{\beta}\left[\frac{3F^2_{\alpha\beta}  (\alpha +\beta -1)^2 \left(\alpha ^3+\beta ^3\right)}{393216 \pi ^8 \alpha ^4 \beta ^4}+\frac{ F_{\alpha\beta}^2 }{131072 \pi ^8 \alpha ^2 \beta ^2}\right.\\
		& &\left.+\frac{F_{\alpha\beta}(\alpha +\beta ) (\alpha +\beta -1)^2 m_Q^2 \left(\alpha ^3+\beta ^3\right)}{196608 \pi ^8 \alpha ^4 \beta ^4}\right],\\
		\rho^{\expval{qG\bar{q}}^2}&=&\expval{qG\bar{q}}^2\int_{\alpha}\left[ \frac{H_{\alpha}^2 }{8192 \pi ^6 (1-\alpha ) \alpha }-\int_{\beta}\frac{F_{\alpha\beta}m_Q^2 }{4096 \pi ^6 \alpha  \beta} \right],\\
		\rho^{\expval{q\bar{q}}^2\expval{G^2}}&=&-\expval{q\bar{q}}^2\expval{G^2}\int_{\alpha}\int_{\beta}\left[\frac{(\alpha +\beta -1)m_Q^4 \left(\alpha ^4+\alpha ^3 (\beta -1)+\alpha  \beta ^3+(\beta -1) \beta ^3\right)}{36864 \pi ^6 \alpha ^3 \beta ^3} \right.\\
		& &\left. +\frac{ (\alpha +\beta -1) F_{\alpha\beta} m_Q^2 \left( 5\alpha ^3+3 \alpha ^2 (\beta -1)+3 \alpha  \beta ^2+(5\beta -3) \beta ^2\right)}{36864 \pi ^6 \alpha ^3 \beta ^3}\right.\\
		& &\left.-\frac{F_{\alpha\beta}  \left(F_{\alpha\beta}-2 \alpha  m_Q^2\right)}{24576 \pi ^6 \alpha ^2 \beta }   \right],\\
		\rho^{\expval{q\bar{q}}^2\expval{qG\bar{q}}}&=&-\expval{q\bar{q}}^2\expval{qG\bar{q}}\int_{\alpha}\left[\frac{H_{\alpha} m_Q}{256 \pi ^4 (1-\alpha )}+\frac{H_{\alpha} m_Q }{256 \pi ^4 \alpha }\right],\\
		\rho^{\expval{q\bar{q}}^4}&=&\int_{\alpha}\frac{m_Q^2 \expval{q\bar{q}}^4}{144 \pi ^2}.
	\end{eqnarray}
	
	\subsection{$0^+$ $\Lambda_Q\bar{\Sigma}_Q$ States}
	\begin{eqnarray}
		\rho^{\text{pert}}&=& - \int_{\alpha}\int_{\beta} \frac{F_{\alpha\beta}^7 (\alpha +\beta -1)^4}{55050240 \pi ^{10} \alpha ^6 \beta ^6},\\	
		\rho^{\expval{q\bar{q}}}&=&\int_{\alpha}\int_{\beta}\frac{F_{\alpha\beta}^5 m_Q \expval{q\bar{q}} (\alpha -\beta ) (\alpha +\beta -1)^3}{245760 \pi ^8 \alpha ^5 \beta ^5},\\
		\rho^{\expval{G^2}}&=& -\expval{g_s^2G^2}\int_{\alpha}\int_{\beta}\left[\frac{F_{\alpha\beta}^5 (2 \alpha -\beta +1) (\alpha +\beta -1)^2}{15728640 \pi ^{10} \alpha ^5 \beta ^4}\right.\\
		& &\left.+\frac{F_{\alpha\beta}^4 m_Q^2  (\alpha +\beta -1)^4 \left(\alpha ^3+\beta ^3\right)}{18874368 \pi ^{10} \alpha ^6 \beta ^6}\right],\\
		\rho^{\expval{qG\bar{q}}}&=&-m_Q\expval{qG\bar{q}}\int_{\alpha}\int_{\beta}\left[\frac{F_{\alpha\beta}^4  (\alpha -\beta ) (\alpha +\beta -1)^2}{65536 \pi ^8 \alpha ^4 \beta ^4}-\frac{F_{\alpha\beta}^4  (\alpha +\beta -1)^3}{196608 \pi ^8 \alpha ^5 \beta ^3}\right],\\
		\rho^{\expval{q\bar{q}}^2}&=&-\expval{q\bar{q}}^2\int_{\alpha}\int_{\beta}\frac{F_{\alpha\beta}^3 (\alpha +\beta -1) \left(F_{\alpha\beta}-2 m_Q^2 (\alpha +\beta -1)\right)}{6144 \pi ^6 \alpha ^3 \beta ^3},\\
		\rho^{\expval{G^3}}&=&-\expval{g_s^3G^3}\int_{\alpha}\int_{\beta}\frac{F_{\alpha\beta}^3 (\alpha +\beta -1)^4 \left(F_{\alpha\beta} \left(\alpha ^3+\beta ^3\right)+8 m_Q^2 \left(\alpha ^4+\beta ^4\right)\right)}{75497472 \pi ^{10} \alpha ^6 \beta ^6},\\
		\rho^{\expval{q\bar{q}}\expval{G^2}}&=&m_Q\expval{q\bar{q}}\expval{g_s^2G^2}\int_{\alpha}\int_{\beta}\left[\frac{F_{\alpha\beta}^3  (\alpha -3 \beta +1) (\alpha +\beta -1)}{196608 \pi ^8 \alpha ^3 \beta ^3}\right.\\
		& &\left.+\frac{F_{\alpha\beta}^2 (\alpha-\beta) (\alpha +\beta -1)^3 \left(F_{\alpha\beta} \left(\alpha ^2+\alpha  \beta +\beta ^2\right)+m_Q^2 \left(\alpha ^3+\beta ^3\right)\right)}{294912 \pi ^8 \alpha ^5 \beta ^5}\right],\\
		\rho^{\expval{q\bar{q}}\expval{qG\bar{q}}}&=&\expval{q\bar{q}}\expval{qG\bar{q}}\int_{\alpha}\int_{\beta}\left[\frac{F_{\alpha\beta}^2 m_Q^2  (-\alpha -\beta +1)}{1024 \pi ^6 \alpha ^2 \beta ^2}+\frac{F_{\alpha\beta}^3}{3072 \pi ^6 \alpha ^2 \beta ^2}\right],\\
		\rho^{\expval{q\bar{q}}^3}&=&\int_{\alpha}\int_{\beta}\frac{F_{\alpha\beta}^2 m_Q \expval{q\bar{q}}^3 (\alpha -\beta )}{384 \pi ^4 \alpha ^2 \beta ^2},\\
		\rho^{\expval{qG\bar{q}}\expval{G^2}}&=& m_Q\expval{qG\bar{q}}\expval{G^2}\int_{\alpha}\int_{\beta}\left[-\frac{3F^2_{\alpha\beta} (\alpha -\beta ) (\alpha +\beta -1)^2 \left(\alpha ^2+\alpha  \beta +\beta ^2\right)}{393216 \pi ^8 \alpha ^4 \beta ^4}\right.\\
		& &\left.-\frac{F_{\alpha\beta}(\alpha -\beta ) (\alpha +\beta -1)^2 m_Q^2 \left(\alpha ^3+\beta ^3\right)}{196608 \pi ^8 \alpha ^4 \beta ^4}+\frac{(2 \beta -1) F_{\alpha\beta}^2 }{131072 \pi ^8 \alpha ^2 \beta ^2}\right],\\
		\rho^{\expval{qG\bar{q}}^2}&=&\expval{qG\bar{q}}^2\int_{\alpha}\left[ \frac{H_{\alpha}^2 }{8192 \pi ^6 (1-\alpha ) \alpha }+\int_{\beta}\frac{F_{\alpha\beta}m_Q^2 }{4096 \pi ^6 \alpha  \beta} \right],\\
		\rho^{\expval{q\bar{q}}^2\expval{G^2}}&=&\expval{q\bar{q}}^2\expval{G^2}\int_{\alpha}\int_{\beta}\left[\frac{(\alpha +\beta -1)m_Q^4 \left(\alpha ^4+\alpha ^3 (\beta -1)+\alpha  \beta ^3+(\beta -1) \beta ^3\right)}{36864 \pi ^6 \alpha ^3 \beta ^3}\right.\\
		& &\left.+\frac{ (\alpha +\beta -1) F_{\alpha\beta} m_Q^2 \left( \alpha ^3+3 \alpha ^2 (\beta -1)+3 \alpha  \beta ^2+( \beta -3) \beta ^2\right)}{36864 \pi ^6 \alpha ^3 \beta ^3}    \right.,\\
		& & \left.+\frac{F_{\alpha\beta}  \left(F_{\alpha\beta}+2 \alpha  m_Q^2\right)}{24576 \pi ^6 \alpha ^2 \beta }\right],\\
		\rho^{\expval{q\bar{q}}^4}&=&-\int_{\alpha}\frac{m_Q^2 \expval{q\bar{q}}^4}{144 \pi ^2}.
	\end{eqnarray}
	
	\subsection{$1^-$ $\Lambda_Q\bar{\Sigma}_Q$ States}
	\begin{eqnarray}
		\rho^{\text{pert}}&=& - \int_{\alpha}\int_{\beta} \frac{F_{\alpha\beta}^7 (\alpha +\beta -1)^4(\alpha +\beta +4)}{275251200  \pi ^{10} \alpha ^6 \beta ^6},\\	
		\rho^{\expval{q\bar{q}}}&=&-\int_{\alpha}\int_{\beta}\frac{F_{\alpha\beta}^5 m_Q \expval{q\bar{q}} (\alpha  (\beta +4)+(\beta +3) \beta) (\alpha +\beta -1)^3}{983040 \pi ^8 \alpha ^5 \beta ^5},\\
		\rho^{\expval{G^2}}&=& -\expval{g_s^2G^2}\int_{\alpha}\int_{\beta}\left[\frac{F_{\alpha\beta}^5 (3 \alpha ^2+2 \alpha  (\beta +3)-\beta ^2-2 \beta +3) (\alpha +\beta -1)^2}{62914560 \pi ^{10} \alpha ^5 \beta ^4}\right.\\
		& &\left.+\frac{F_{\alpha\beta}^4 m_Q^2  (\alpha +\beta -1)^4 (\alpha ^4+\alpha ^3 (\beta +4)+\alpha  \beta ^3+(\beta +4) \beta ^3)}{94371840 \pi ^{10} \alpha ^6 \beta ^6}\right],\\
		\rho^{\expval{qG\bar{q}}}&=&m_Q\expval{qG\bar{q}}\int_{\alpha}\int_{\beta}\left[\frac{F_{\alpha\beta}^4  (\alpha +\beta -1)^2(\alpha  (\beta +3)+(\beta +2) \beta)}{196608 \pi ^8 \alpha ^4 \beta ^4}\right.\\
		& &\left.+\frac{F_{\alpha\beta}^4 ( \alpha +\beta +3)(\alpha +\beta -1)^3}{786432 \pi ^8 \alpha ^5 \beta ^3}\right],\\
		\rho^{\expval{q\bar{q}}^2}&=&-\expval{q\bar{q}}^2\int_{\alpha}\int_{\beta}\frac{F_{\alpha\beta}^3 (\alpha +\beta -1) \left(F_{\alpha\beta}(\alpha +\beta +1)+4 m_Q^2 (\alpha +\beta -1)\right)}{12288 \pi ^6 \alpha ^3 \beta ^3},\\
		\rho^{\expval{G^3}}&=&-\expval{g_s^3G^3}\int_{\alpha}\int_{\beta}\frac{F_{\alpha\beta}^3 (\alpha +\beta -1)^4(\alpha+\beta+4) \left(F_{\alpha\beta} \left(\alpha ^3+\beta ^3\right)+8 m_Q^2 \left(\alpha ^4+\beta ^4\right)\right)}{377487360 \pi ^{10} \alpha ^6 \beta ^6},\\
		\rho^{\expval{q\bar{q}}\expval{G^2}}&=&-m_Q\expval{q\bar{q}}\expval{g_s^2G^2}\int_{\alpha}\int_{\beta}\left[\frac{F_{\alpha\beta}^3  (\alpha + \beta -1) (\alpha  \beta +\alpha +\beta ^2+1)}{196608 \pi ^8 \alpha ^3 \beta ^3}\right.\\
		& & \left.+\frac{F_{\alpha\beta}^3 (\alpha +\beta -1)^3(4 \alpha ^3+\alpha  \beta ^3+(\beta +3) \beta ^3) }{1179648 \pi ^8 \alpha ^5 \beta ^5}\right.\\
		& &\left.+\frac{F_{\alpha\beta}^2 m_Q^2 (\alpha ^4 (\beta +4)+\alpha ^3 (\beta +3) \beta +\alpha  (\beta +4) \beta ^3+(\beta +3) \beta ^4)}{1179648 \pi ^8 \alpha ^5 \beta ^5}\right],\\
		\rho^{\expval{q\bar{q}}\expval{qG\bar{q}}}&=&\expval{q\bar{q}}\expval{qG\bar{q}}\int_{\alpha}\int_{\beta}\left[-\frac{F_{\alpha\beta}^2 m_Q^2  (-\alpha -\beta +1)}{1024 \pi ^6 \alpha ^2 \beta ^2}-\frac{F_{\alpha\beta}^3(-\alpha -\beta +1)}{3072 \pi ^6 \alpha ^2 \beta ^2}+\frac{F_{\alpha\beta}^3}{3072 \pi ^6 \alpha ^2 \beta ^2}\right],\\
		\rho^{\expval{q\bar{q}}^3}&=&-\int_{\alpha}\int_{\beta}\frac{F_{\alpha\beta}^2 m_Q \expval{q\bar{q}}^3 (\alpha  (\beta +1)+\beta ^2)}{384 \pi ^4 \alpha ^2 \beta ^2},\\
		\rho^{\expval{qG\bar{q}}\expval{G^2}}&=& m_Q\expval{qG\bar{q}}\expval{G^2}\int_{\alpha}\int_{\beta}\left[\frac{3F^2_{\alpha\beta} (3 \alpha ^3+\alpha  \beta ^3+(\beta +2) \beta ^3 ) (\alpha +\beta -1)^2 }{1179648  \pi ^8 \alpha ^4 \beta ^4}\right.\\
		& &\left.+\frac{F_{\alpha\beta} (\alpha +\beta -1)^2 m_Q^2 \left(\alpha ^4 (\beta +3)+\alpha ^3 (\beta +2) \beta +\alpha  (\beta +3) \beta ^3+(\beta +2) \beta ^4\right)}{589824 \pi ^8 \alpha ^4 \beta ^4}\right.\\
		& &\left.-\frac{(\alpha  \beta +\beta ^2-\beta +1) F_{\alpha\beta}^2 }{131072 \pi ^8 \alpha ^2 \beta ^2}\right],\\
		\rho^{\expval{qG\bar{q}}^2}&=&\expval{qG\bar{q}}^2\int_{\alpha}\left\{\frac{H_{\alpha}^2 }{8192 \pi ^6 (1-\alpha ) \alpha }-\int_{\beta}\left[\frac{F_{\alpha\beta}^2}{8192\pi ^6 \alpha  \beta}+\frac{F_{\alpha\beta}m_Q^2 }{4096 \pi ^6 \alpha  \beta} \right]\right\},\\
		\rho^{\expval{q\bar{q}}^2\expval{G^2}}&=&-\expval{q\bar{q}}^2\expval{G^2}\int_{\alpha}\int_{\beta}\left[\frac{(\alpha +\beta -1)m_Q^4 \left(\alpha ^4+\alpha ^3 (\beta -1)+\alpha  \beta ^3+(\beta -1) \beta ^3\right)}{36864 \pi ^6 \alpha ^3 \beta ^3}  \right.\\
		& +& \left.\frac{ (\alpha +\beta -1) F_{\alpha\beta} m_Q^2 \left(\alpha ^4+\alpha ^3 (\beta +4)+3 \alpha ^2 (\beta -1)+\alpha  (\beta +3) \beta ^2+\left(\beta ^2+4 \beta -3\right) \beta ^2\right)}{36864 \pi ^6 \alpha ^3 \beta ^3}\right.\\
		& &\left.\frac{F_{\alpha\beta}  \left(F_{\alpha\beta}(\alpha+\beta)+2 \alpha  m_Q^2\right)}{24576 \pi ^6 \alpha ^2 \beta }\right],\\
		\rho^{\expval{q\bar{q}}^2\expval{qG\bar{q}}}&=&\expval{q\bar{q}}^2\expval{qG\bar{q}}\int_{\alpha}\left[-\frac{H_{\alpha} m_Q }{256 \pi ^4 (1-\alpha )}-\frac{H_{\alpha} m_Q }{256 \pi ^4 \alpha }+\int_{\beta}\frac{F_{\alpha\beta} m_Q}{256 \pi ^4 \alpha }\right],\\
		\rho^{\expval{q\bar{q}}^4}&=&\int_{\alpha}\frac{m_Q^2 \expval{q\bar{q}}^4}{144 \pi ^2}.
	\end{eqnarray}
	
	\subsection{$1^+$ $\Lambda_Q\bar{\Sigma}_Q$ States}
	\begin{eqnarray}
		\rho^{\text{pert}}&=& - \int_{\alpha}\int_{\beta} \frac{F_{\alpha\beta}^7 (\alpha +\beta -1)^4(\alpha +\beta +4)}{275251200  \pi ^{10} \alpha ^6 \beta ^6},\\	
		\rho^{\expval{q\bar{q}}}&=&-\int_{\alpha}\int_{\beta}\frac{F_{\alpha\beta}^5 m_Q \expval{q\bar{q}} (\alpha  (\beta -4)+(\beta +3) \beta) (\alpha +\beta -1)^3}{983040 \pi ^8 \alpha ^5 \beta ^5},\\
		\rho^{\expval{G^2}}&=& -\expval{g_s^2G^2}\int_{\alpha}\int_{\beta}\left[\frac{F_{\alpha\beta}^5 (3 \alpha ^2+2 \alpha  (\beta +3)-\beta ^2-2 \beta +3) (\alpha +\beta -1)^2}{62914560 \pi ^{10} \alpha ^5 \beta ^4}\right.\\
		& &\left.+\frac{F_{\alpha\beta}^4 m_Q^2  (\alpha +\beta -1)^4 (\alpha ^4+\alpha ^3 (\beta +4)+\alpha  \beta ^3+(\beta +4) \beta ^3)}{94371840 \pi ^{10} \alpha ^6 \beta ^6}\right],\\
		\rho^{\expval{qG\bar{q}}}&=&m_Q\expval{qG\bar{q}}\int_{\alpha}\int_{\beta}\left[\frac{F_{\alpha\beta}^4  (\alpha +\beta -1)^2(\alpha  (\beta -3)+(\beta +2) \beta)}{196608 \pi ^8 \alpha ^4 \beta ^4}\right.\\
		& &\left.+\frac{F_{\alpha\beta}^4 ( \alpha +\beta +3)(\alpha +\beta -1)^3}{786432 \pi ^8 \alpha ^5 \beta ^3}\right],\\
		\rho^{\expval{q\bar{q}}^2}&=&-\expval{q\bar{q}}^2\int_{\alpha}\int_{\beta}\frac{F_{\alpha\beta}^3 (\alpha +\beta -1) \left(F_{\alpha\beta}(\alpha +\beta +1)-4 m_Q^2 (\alpha +\beta -1)\right)}{12288 \pi ^6 \alpha ^3 \beta ^3},\\
		\rho^{\expval{G^3}}&=&-\expval{g_s^3G^3}\int_{\alpha}\int_{\beta}\frac{F_{\alpha\beta}^3 (\alpha +\beta -1)^4(\alpha +\beta +4) \left(F_{\alpha\beta} \left(\alpha ^3+\beta ^3\right)+8 m_Q^2 \left(\alpha ^4+\beta ^4\right)\right)}{75497472 \pi ^{10} \alpha ^6 \beta ^6},\\
		\rho^{\expval{q\bar{q}}\expval{G^2}}&=&-m_Q\expval{q\bar{q}}\expval{g_s^2G^2}\int_{\alpha}\int_{\beta}\left[\frac{F_{\alpha\beta}^3  (\alpha + \beta -1) (\alpha  (\beta -1)+\beta ^2+2 \beta -1)}{196608 \pi ^8 \alpha ^3 \beta ^3}\right.\\
		& & \left.+\frac{F_{\alpha\beta}^3 (\alpha +\beta -1)^3(-4 \alpha ^3+\alpha  \beta ^3+(\beta +3) \beta ^3) }{1179648 \pi ^8 \alpha ^5 \beta ^5}\right.\\
		& &\left.+\frac{F_{\alpha\beta}^2 m_Q^2 (\alpha ^4 (\beta -4)+\alpha ^3 (\beta +3) \beta +\alpha  (\beta -4) \beta ^3+(\beta +3) \beta ^4)}{1179648 \pi ^8 \alpha ^5 \beta ^5}\right],\\
		\rho^{\expval{q\bar{q}}\expval{qG\bar{q}}}&=&\expval{q\bar{q}}\expval{qG\bar{q}}\int_{\alpha}\int_{\beta}\left[\frac{F_{\alpha\beta}^2 m_Q^2  (-\alpha -\beta +1)}{1024 \pi ^6 \alpha ^2 \beta ^2}-\frac{F_{\alpha\beta}^3(-\alpha -\beta +1)}{3072 \pi ^6 \alpha ^2 \beta ^2}+\frac{F_{\alpha\beta}^3}{3072 \pi ^6 \alpha ^2 \beta ^2}\right],\\
		\rho^{\expval{q\bar{q}}^3}&=&-\int_{\alpha}\int_{\beta}\frac{F_{\alpha\beta}^2 m_Q \expval{q\bar{q}}^3 (\alpha  (\beta -1)+\beta ^2)}{384 \pi ^4 \alpha ^2 \beta ^2},\\
		\rho^{\expval{qG\bar{q}}\expval{G^2}}&=& m_Q\expval{qG\bar{q}}\expval{G^2}\int_{\alpha}\int_{\beta}\left[\frac{3F^2_{\alpha\beta} (-3 \alpha ^3+\alpha  \beta ^3+(\beta +2) \beta ^3 ) (\alpha +\beta -1)^2 }{1179648  \pi ^8 \alpha ^4 \beta ^4}\right.\\
		& &\left.+\frac{F_{\alpha\beta} (\alpha +\beta -1)^2 m_Q^2 \left(\alpha ^4 (\beta -3)+\alpha ^3 (\beta +2) \beta +\alpha  (\beta -3) \beta ^3+(\beta +2) \beta ^4\right)}{589824 \pi ^8 \alpha ^4 \beta ^4}\right.\\
		& &\left.+\frac{(\alpha  \beta +\beta ^2+\beta -1) F_{\alpha\beta}^2 }{131072 \pi ^8 \alpha ^2 \beta ^2}\right],\\
		\rho^{\expval{qG\bar{q}}^2}&=&\expval{qG\bar{q}}^2\int_{\alpha}\left\{\frac{H_{\alpha}^2 }{8192 \pi ^6 (1-\alpha ) \alpha }-\int_{\beta}\left[\frac{F_{\alpha\beta}m_Q^2 }{4096 \pi ^6 \alpha  \beta} +\frac{F_{\alpha\beta}^2}{8192\pi ^6 \alpha  \beta}\right]\right\},\\
		\rho^{\expval{q\bar{q}}^2\expval{G^2}}&=&\expval{q\bar{q}}^2\expval{G^2}\int_{\alpha}\int_{\beta}\left[\frac{(\alpha +\beta -1)m_Q^4 \left(\alpha ^4+\alpha ^3 (\beta -1)+\alpha  \beta ^3+(\beta -1) \beta ^3\right)}{36864 \pi ^6 \alpha ^3 \beta ^3}  \right.\\
		&-&  \left.\frac{ (\alpha +\beta -1) F_{\alpha\beta} m_Q^2 \left(\alpha ^4+\alpha ^3 (\beta -2)-3 \alpha ^2 (\beta -1)+\alpha  (\beta -3) \beta ^2+\left(\beta ^2-2 \beta +3\right) \beta ^2\right)}{36864 \pi ^6 \alpha ^3 \beta ^3}\right.\\
		& &+\left.\frac{F_{\alpha\beta}  \left(F_{\alpha\beta}(\alpha+\beta)+2 \alpha  m_Q^2\right)}{24576 \pi ^6 \alpha ^2 \beta }\right],\\
		\rho^{\expval{q\bar{q}}^2\expval{qG\bar{q}}}&=&\expval{q\bar{q}}^2\expval{qG\bar{q}}\int_{\alpha}\int_{\beta}\frac{F_{\alpha\beta} m_Q}{256 \pi ^4 \alpha },\\
		\rho^{\expval{q\bar{q}}^4}&=&-\int_{\alpha}\frac{m_Q^2 \expval{q\bar{q}}^4}{144 \pi ^2}.
	\end{eqnarray}

\end{appendix}


\end{document}